\documentclass[10pt,twocolumn,journal]{IEEEtran}
\usepackage{latexsym}
\usepackage{float}
\usepackage{amsfonts}
\usepackage{amsbsy}
\usepackage{amssymb}
\usepackage{times}
\usepackage{graphicx}
\usepackage{setspace}
\usepackage{enumerate}
\usepackage[usenames]{color}
\usepackage[dvips]{pstcol}
\usepackage{epstopdf}
\usepackage{cite}
\usepackage{amssymb}
\usepackage{amsfonts}
\usepackage{graphicx}
\usepackage{epsfig}
\usepackage{psfrag}
\usepackage{xcolor}
\usepackage{amsfonts, bm}
\usepackage{epstopdf}
\usepackage{cite}
\usepackage{color}
\usepackage{xcolor}
\usepackage{subfig}
\usepackage{verbatim}
\usepackage{multirow}
\usepackage{array}
\usepackage{booktabs}
\usepackage{amsthm}
\usepackage{makecell}
\usepackage{units}
\usepackage[linesnumbered, ruled]{algorithm2e}
\usepackage{algpseudocode}
\usepackage[ruled]{algorithm2e}
\usepackage{amsmath}

\IEEEoverridecommandlockouts
\begin{document}
\title{Feature Allocation for Semantic Communication with Space-Time Importance Awareness}
\author{Kequan Zhou, Guangyi Zhang, Yunlong Cai, Qiyu Hu, Guanding Yu, and A. Lee Swindlehurst
	\thanks{ K. Zhou, G. Zhang, Y. Cai, Q. Hu, and G. Yu are with the College of Information Science and Electronic Engineering, Zhejiang University, Hangzhou 310027, China (e-mail: kqzhou@zju.edu.cn; zhangguangyi@zju.edu.cn; qiyhu@zju.edu.cn; ylcai@zju.edu.cn; yuguanding@zju.edu.cn).
	
	A. Lee Swindlehurst is with the Center for Pervasive Communications and Computing, University of California at Irvine, Irvine, CA 92697 USA (e-mail: swindle@uci.edu).} }

\maketitle
\vspace{-3.3em}
\begin{abstract}
In the realm of semantic communication, the significance of encoded features can vary, while wireless channels are known to exhibit fluctuations across multiple subchannels in different domains.
Consequently, critical features may traverse subchannels with poor states, resulting in performance degradation.
To tackle this challenge, we introduce a framework called Feature Allocation for Semantic Transmission (FAST), which offers adaptability to channel fluctuations across both spatial and temporal domains.
In particular, an importance evaluator is first developed to assess the importance of various features.
In the temporal domain, channel prediction is utilized to estimate future channel state information (CSI).
Subsequently, feature allocation is implemented by assigning suitable transmission time slots to different features.
Furthermore, we extend FAST to the space-time domain, considering two common scenarios: precoding-free and precoding-based multiple-input multiple-output (MIMO) systems.
An important attribute of FAST is its versatility, requiring no intricate fine-tuning.
Simulation results demonstrate that this approach significantly enhances the performance of semantic communication systems in image transmission.
It retains its superiority even when faced with substantial changes in system configuration.
\end{abstract}

\IEEEpeerreviewmaketitle

\section{Introduction}
Recent years have witnessed the emergence of intelligent services such as virtual reality and autonomous driving \cite{Strinati20216g, Yang2023semantic, Shi2021from}.
These services demand higher data rates to manage the influx of massive data traffic.
However, conventional communication systems have reached a bottleneck in meeting these requirements, prompting a paradigm shift from Shannon's legacy.
Semantic communication, a novel paradigm, has recently garnered substantial attention \cite{Niu2022a, Strinati2021toward, Uysal2022semantic, Choi2022a}.
Diverging from traditional communication, semantic communication focuses on transmitting the inherent semantic information within the source data, ultimately enhancing communication efficiency.
\subsection{Prior Work}
The rapid advancement of deep learning (DL) techniques has given rise to numerous semantic communication systems in recent years \cite{Xie2021deep, Bourtsoulatze2019deep, Yan2021deep, Wang2023wireless}.
Unlike the segmented structure of traditional communication, semantic communication jointly designs source and channel coding.
A pioneering DL-based joint source-channel coding scheme known as DeepJSCC was introduced in \cite{Bourtsoulatze2019deep} for image transmission.
Building upon this, various subsequent designs and enhancements have been put forth \cite{Sun2022deep, Zhao2023semantic, Xiao2023wireless, Yue2023learned, Hu2023scalable, Yang2022deep, Du2023yolo, Guo2023semantic, Wang2021performance, Kim2023distributed}.
The notion of semantic importance has been a common theme in these innovations.
In \cite{Sun2022deep}, semantic importance was employed to weight different features, boosting performance in image-related tasks.
For transmission of text data, \cite{Zhao2023semantic} introduced an optimization framework designed to capture feature importance, aimed at transmitting only the most important features.
Additionally, \cite{Xiao2023wireless} utilized a learnable entropy model to estimate the importance of semantic features in speech transmission, leading to minimized end-to-end rate-distortion performance.
Furthermore, a semantic importance-aware approach was proposed in \cite{Wang2021performance}, facilitating the joint optimization of resource allocation and system performance.

In addition to the consideration of semantic importance, recent research has placed an emphasis on incorporating channel state information (CSI) during model training \cite{Ding2021snr, Xu2021wireless, Bao2021adjscc, Wang2023learn, Wu2022channel, Zhang2022unified}.
For instance, a novel SNR-adaptive system was introduced in \cite{Xu2021wireless}, leveraging attention mechanisms and joint training with CSI to operate effectively across varying SNRs.
Building on this, \cite{Bao2021adjscc} proposed an attention-based multi-layer joint source-channel coding (JSCC) architecture for progressive image transmission.
A task-attention model capable of adapting to new channel distributions through the use of a few pilot blocks in new environments was developed in \cite{Wang2023learn}.
Additionally, \cite{Wu2022channel} integrated a dual attention mechanism with classical orthogonal frequency division multiplexing (OFDM) techniques to combat fading.

Beyond the aforementioned works focusing on single-input single-output (SISO) systems, there is a growing interest in semantic communication over multiple-input multiple-output (MIMO) channels \cite{O2017deep, Song2020benchmarking, Zhang2022svd, Wu2023deep, Zhang2023scan, Shi2023excess, Yao2022versatile, Yang2022ofdm}.
Specifically, the exploration of DL-based MIMO communication systems was initiated in \cite{O2017deep}, featuring an autoencoder architecture combined with classical Singular Value Decomposition (SVD) precoding.
To enhance the channel adaptability of semantic communication systems over MIMO channels, \cite{Wu2023deep} introduced a vision Transformer (ViT)-based DeepJSCC scheme, harnessing the self-attention mechanism of ViT.
Moreover, \cite{Yang2022ofdm} devised a JSCC model jointly trained with explicit OFDM blocks, incorporating adversarial training techniques and a novel power allocation scheme to combat multipath fading.
\subsection{Motivation and Contributions}
Although previous research has made significant strides, two limitations impede the practical use of these techniques.

\begin{itemize}
	\item [\textbf{P1.}] The models in the work above are typically trained using additive
	white Gaussian noise (AWGN) channels and operate at a fixed SNR, failing to fully capture the dynamic characteristics of real-world wireless channels. Consequently, their performance may be suboptimal when faced with changing channel conditions.
	\item [\textbf{P2.}] The performance of the models is constrained by system-specific configurations, such as the number of antennas.
	When the configurations change, the models inevitably experience performance degradation.
	Adapting the models to new configurations through parameter fine-tuning becomes necessary to maintain high system performance.
	However, this process incurs substantial computational overhead, which may not be feasible in real-world scenarios.
\end{itemize}

\begin{figure}[htb]
	\begin{centering}
		\includegraphics[width=0.41\textwidth]{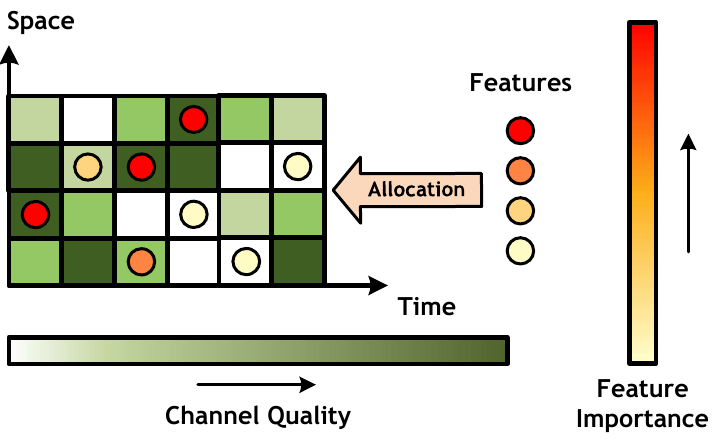}
		\par\end{centering}
	\captionsetup{font=footnotesize}
	\caption{The illustration of feature allocation.}
	\label{STAlloc}
\end{figure}

To address these challenges, we propose a versatile framework called Feature Allocation for Semantic Transmission (FAST), focusing on preserving crucial features during transmission to enhance overall system performance.
Semantic signals contain distinct information elements, each of different importance for the given task \cite{Sun2022deep}.
This contrasts with the independent and identically distributed (i.i.d.) signals normally assumed in traditional communication systems.
Additionally, the physical channel fluctuates across multiple dimensions, including different time slots and antennas, occasionally experiencing fading.
As illustrated in Fig. \ref{STAlloc}, a simple approach involving random allocation of features to various subchannels can make important features subject to unexpected fading, corrupting them and causing performance degradation.
Hence, improved performance can be achieved by strategically assigning favorable subchannels to important features while allowing less critical features to traverse less ideal subchannels.

To implement FAST, it is essential to evaluate the importance of each feature and the quality of each subchannel in advance at the transmitter.
For this purpose, an importance evaluator is first developed to calculate feature importance.
The evaluator's design relies on a gradient-based algorithm and the knowledge distillation technique.
To assess the quality of subchannels in various domains, we initially address the time domain.
In this context, we consider a SISO system and perform channel prediction at the receiver to estimate future CSI, which is subsequently fed back to the transmitter.
Each subchannel is a time slot, whose quality is evaluated based on the gain of the CSI for that time slot.
Next, we proceed to design the time domain feature allocation.

Driven by the prevalence of MIMO systems in modern communication, we extend FAST to the spatial domain.
In practice, there are two scenarios that are typically considered: precoding-free and precoding-based systems, both of which are thoroughly explored in this paper.
We refer to each subchannel as a space-time resource block, with its quality determined on a block-by-block basis.
In precoding-free MIMO systems, we employ the minimum mean square error (MMSE) equalization technique at the receiver.
Here, a resource block corresponds to a transmit antenna operating within a specific time slot, and its quality is determined by evaluating the received signal-to-interference-plus-noise ratio (SINR).
For precoding-based MIMO systems, we utilize the classical SVD precoding technique.
In this case, a resource block represents a specific parallel subchannel within a particular time slot, and its quality is assessed based on the corresponding singular value.
Subsequently, a space-time feature allocator is designed following the concept of feature allocation depicted in Fig. \ref{STAlloc}.

\begin{figure*}[t]
	\centering
	\subfloat[Components of the basic model.]{\includegraphics[width=0.78\textwidth]{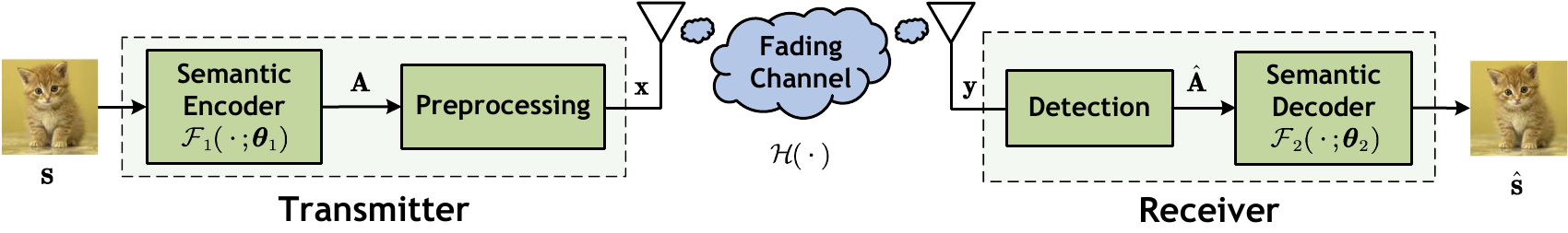}\label{FrameworkBasic}}
	
	\subfloat[Components of FAST.]{\includegraphics[width=0.99\textwidth]{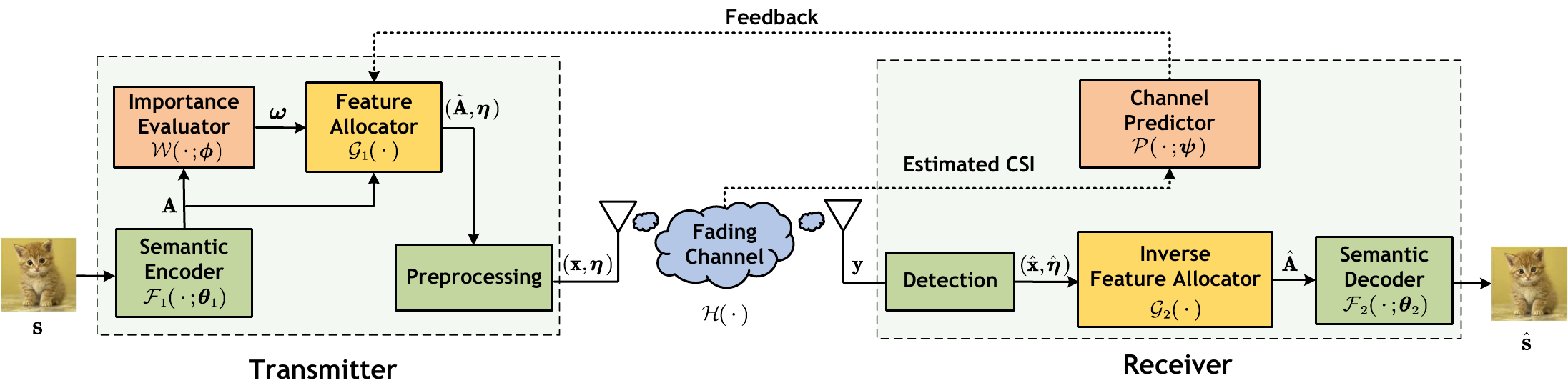}\label{FrameworkFast}}
	\captionsetup{font=footnotesize}
	\caption{The framework of the proposed FAST approach.}
	\label{Framework}
\end{figure*}

The proposed method enhances the performance of semantic communication systems by safeguarding essential features, effectively addressing \textbf{P1}.
Moreover, it constitutes a versatile design that demands no fine-tuning and facilitates straightforward deployment, ensuring that performance improvements are retained even in the face of system configuration changes, thus positioning it as a promising solution to \textbf{P2}.
Furthermore, the acronym FAST embodies not only its adaptability to changing system configurations but also the lower latency and improved efficiency of semantic communication.
The key contributions of this paper are summarized as follows.

\begin{itemize}
\item We introduce FAST, a multi-dimensional framework that enhances the performance of semantic communication systems.
This method provides a versatile solution, necessitating no fine-tuning and delivering resilient performance, even in the presence of dynamic system configuration changes.
\item We devise an importance evaluator, an advanced tool at the transmit end, which employs an evaluation algorithm in conjunction with the knowledge distillation technique to calculate feature importance.
\item We create an innovative space-time feature allocator and design distinct allocation schemes to cater to both precoding-free and precoding-based MIMO systems.
\item Simulation results demonstrate that the proposed FAST approach yields significant performance gains compared to semantic communication systems lacking intelligent feature resource allocation.
\end{itemize}

\subsection{Organization and Notation}
The rest of the paper is organized as follows.
Section \ref{SecFramework} introduces the FAST framework.
The details of the FAST approach are presented in Section \ref{SecTimeAlloc}.
Then, Section \ref{SecSpaceTimeAlloc} discusses the design of the space-time feature allocation.
Simulation results are provided in Section \ref{Simulation}, and Section \ref{Conclusion} concludes the paper.

Scalars, vectors, and matrices are respectively denoted by lower case, boldface lower case, and boldface upper case letters.
For a matrix $\mathbf{A}$, $\mathbf{A}^H$ is its conjugate transpose.
For a vector $\mathbf{a}$, $\mathbf{a}^T$ and $||\mathbf{a}||$ are its transpose and Euclidean norm, respectively.
Symbols marked with a tilde (e.g., $\tilde{\mathbf{h}}$ and $\tilde{\mathbf{A}}$) denote signals obtained through prediction or resource allocation operations.
Symbols marked with a hat (e.g., $\hat{\mathbf{s}}$ and $\hat{\mathbf{A}}$) represent signals obtained through reconstruction.
Finally, $\mathbb{C}^{m\times n}(\mathbb{R}^{m\times n})$ is the space of $m\times n$ complex (real) matrices.

\section{Framework of FAST} \label{SecFramework}
In this section, we introduce the FAST framework and examine its application to the domain of image transmission.

\begin{figure*}[t]
	\centering
	\includegraphics[width=0.7\textwidth]{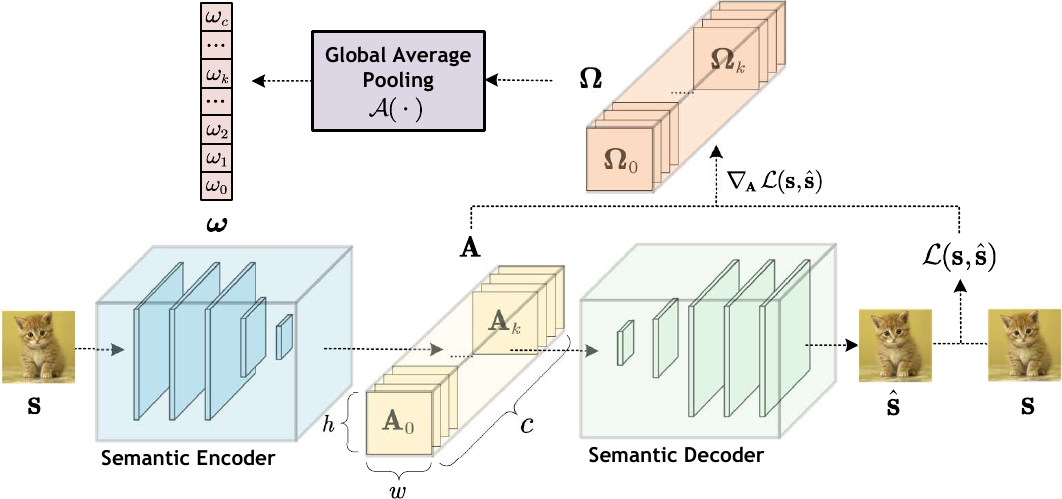}
	\captionsetup{font=footnotesize}
	\caption{The calculation of feature importance based on gradients.}
	\label{Importance}
\end{figure*}

\subsection{Basic Model}
As illustrated in Fig. \ref{Framework}(a), the basic model consists of a pair of semantic encoder and decoder.
An input image is represented by a vector $\mathbf{s}\in \mathbb{R}^l$, and the transmitter first encodes $\mathbf{s}$ into a feature tensor $\mathbf{A}\in \mathbb{R}^{c\times h\times w}$, where $c$ is the number of features and $h\times w$ is the shape of each feature.
The process is represented as
\begin{equation}
	\mathbf{A}=\mathcal{F}_1(\mathbf{s};\bm{\theta}_1),
\end{equation}
where $\bm{\theta}_1$ denotes the parameter set of the encoder $\mathcal{F}_1(\cdot)$.
Afterwards, $\mathbf{A}$ is mapped into the channel input symbols $\mathbf{x}\in \mathbb{C}^k$, where $k$ is the number of symbols.
Moreover, $\mathbf{x}$ is subject to the average power constraint $P$, i.e.,
$\frac{1}{k}||\mathbf{x}||^2\leq P$.

Then, the signal received at the receiver is given by
\begin{equation}
	\mathbf{y}=\mathcal{H}(\mathbf{x};\bm{\Gamma}, \mathbf{n}),
\end{equation}
where $\mathcal{H}(\cdot)$ denotes the transmission process, $\bm{\Gamma}$ is a CSI sequence generated from the channel model, which consists of the CSI values during a period of $t$ time slots, and $\mathbf{n}$ is AWGN sampled from the distribution $\mathcal{CN}(0, \sigma^2\mathbf{I})$.
Note that $\bm{\Gamma}$ is a general representation of CSI sequences.
Particularly, it could be $\mathbf{h}\in \mathbb{C}^{t}$ in SISO scenarios or $\mathbf{H}\in \mathbb{C}^{t\times N_r\times N_t}$ in MIMO scenarios, where $N_r$ and $N_t$ denote the number of receive and transmit antennas, respectively.

Finally, the receiver decodes the features detected from $\mathbf{y}$ to reconstruct the source data, given by
\begin{equation}
	\hat{\mathbf{s}}=\mathcal{F}_2(\hat{\mathbf{A}};\bm{\theta}_2),
\end{equation}
where $\hat{\mathbf{A}}$ and $\bm{\theta}_2$ denote the detected features and the parameter set of the decoder $\mathcal{F}_2(\cdot)$.
In order to jointly train the basic model, i.e., the semantic encoder and decoder, the mean squared error (MSE) between the original image $\mathbf{s}$ and the reconstructed image $\hat{\mathbf{s}}$ is employed as the system loss:
\begin{equation} \label{Loss}
	\mathcal{L}(\hat{\mathbf{s}}, \mathbf{s}) =\frac{1}{l}||\hat{\mathbf{s}}-\mathbf{s}||^2.
\end{equation}
The detailed training procedure is presented in Algorithm \ref{ModelTrain} below.
\subsection{Overview of FAST}
As illustrated in Fig. \ref{Framework}(b), FAST comprises several essential components beyond the basic model, including an importance evaluator, a channel predictor, and a (inverse) feature allocator.
In the framework of FAST, $\mathbf{A}$ is first fed into the importance evaluator, which calculates the importance of each feature, and leads to an importance distribution vector $\bm{\omega}\in \mathbb{R}^c$.
The process is expressed as
\begin{equation}
	\bm{\omega}=\mathcal{W}(\mathbf{A} ; \bm{\phi}),
\end{equation}
where $\bm{\phi}$ denotes the parameter set of the importance evaluator $\mathcal{W(\cdot)}$.
Meanwhile, the channel predictor at the receiver keeps sampling the CSI and makes predictions accordingly, given by
\begin{equation}
	\tilde{\bm{\Gamma}}=\mathcal{P}(\bm{\Gamma}; \bm{\psi}),
\end{equation}
where $\bm{\Gamma}$, $\tilde{\bm{\Gamma}}$, and $\bm{\psi}$ denote the sampled CSI sequence, the predicted CSI sequence, and the parameter set of the channel predictor $\mathcal{P}(\cdot)$.
Then, the predicted CSI sequence is fed back to the transmitter.

With the assistance of the importance evaluator and the channel predictor, the feature allocator can perform feature allocation by strategically organizing the order of the features, given as
\begin{equation}
	(\tilde{\mathbf{A}}, \bm{\eta}) = \mathcal{G}_1(\mathbf{A}, \bm{\omega}, \tilde{\bm{\Gamma}}),
\end{equation}
where $\tilde{\mathbf{A}}$ denotes the organized features, and $\bm{\eta}\in\mathbb{R}^c$ denotes the new feature order.
Afterwards, $\tilde{\mathbf{A}}$ is mapped into the channel input symbols $\mathbf{x}\in \mathbb{C}^k$.
The feature order $\bm{\eta}$ is transmitted as side information along with $\mathbf{x}$ at the cost of several extra bits.

The received signal is recovered as $(\hat{\mathbf{x}}, \hat{\bm{\eta}})$ through detection, and the feature order is restored by the inverse feature allocator $\mathcal{G}_2(\cdot)$, given as
\begin{equation}
	\hat{\mathbf{A}}=\mathcal{G}_2(\hat{\mathbf{x}}, \hat{\bm{\eta}}).
\end{equation}
Finally, the receiver decodes the restored features to reconstruct the source data.

\begin{algorithm}[htb]
	\begin{normalsize}
		\caption{Training algorithm for the basic model} 
		\label{ModelTrain}
		\DontPrintSemicolon
		\KwIn{The training epochs $N$, batch size $B$, training dataset $\mathcal{S}$, and AWGN variance $\sigma^2$.}
		\KwOut{The optimized parameter sets of the semantic encoder and decoder, i.e., $\bm{\theta^*_1}$ and $\bm{\theta^*_2}$.}
		Divide dataset $\mathcal{S}$ into $Q$ batches of data, $\mathbf{S}_1, \mathbf{S}_2, ..., \mathbf{S}_Q$, each with batch size $B$.\\
		\For{$i\leftarrow 1$ \KwTo $N$}{
			\For{$j\leftarrow 1$ \KwTo $Q$}{
			Sample a batch of data, $\mathbf{S}_j = [\mathbf{s}_{j, 1}, \mathbf{s}_{j, 2}, ..., \mathbf{s}_{j, B}]$.\\
			Generate a batch of CSI sequences, $\bm{\Gamma}_j = [\bm{\Gamma}_{j, 1}, \bm{\Gamma}_{j, 2}, ..., \bm{\Gamma}_{j, B}]$, using a given channel model.\\
			Generate a batch of AWGN samples, $\mathbf{N}_j = [\mathbf{n}_{j, 1}, \mathbf{n}_{j, 2}, ..., \mathbf{n}_{j, B}]$, with variance $\sigma^2$.\\
			Compute the encoded symbols $\mathbf{X}_j = \mathcal{F}_1(\mathbf{S}_j; \bm{\theta}_1)$.\\
			Compute the received symbols $\mathbf{Y}_j = \mathcal{H}(\mathbf{X}_j; \bm{\Gamma}_j, \mathbf{N}_j)$.\\
			Compute the reconstructed data $\hat{\mathbf{S}}_j = \mathcal{F}_2(\mathbf{Y}_j; \bm{\theta}_2)$.\\
			Calculate the average loss based on (\ref{Loss}).\\
			Update $\bm{\theta}_1$ and $\bm{\theta}_2$.
			}
		}
	\end{normalsize}
\end{algorithm}

\section{Feature Allocation}  \label{SecTimeAlloc}
In this section, we commence by delving into feature allocation in the time domain and elucidate its implementation details.

\subsection{Importance Evaluator}
The importance evaluator is designed to compute feature importance before the allocation process.

\subsubsection{Feature Importance}
In a typical semantic communication system, various semantic features differ in significance with respect to the overall system performance.
Semantic importance is characterized by the correlation between semantic features and the target task.
To gauge the importance of these semantic features, we employ an algorithm based on neural network interpretability \cite{Selvaraju2017grad}.
This algorithm calculates feature importance by analyzing the gradients of the system loss $\mathcal{L}(\mathbf{s}, \hat{\mathbf{s}})$ with respect to the individual features.
Essentially, the loss reflects the system's performance, and the gradients provide insights into how the loss responds to variations in a particular feature.
This, in turn, reveals the extent of the feature's influence on the overall system performance.

As shown in Fig. \ref{Importance}, we first compute the gradients of $\mathcal{L}(\mathbf{s}, \hat{\mathbf{s}})$ with respect to the $k$-th feature $\mathbf{A}_k\in \mathbb{R}^{h\times w}$ of the feature tensor $\mathbf{A}$, and obtain a gradient matrix, given by
\begin{equation}
	\bm{\Omega}_k=\nabla_{\mathbf{A}_k}\mathcal{L}(\mathbf{s}, \hat{\mathbf{s}})=[\frac{\partial{\mathcal{L}(\mathbf{s}, \hat{\mathbf{s}})}}{\partial{a_{k,ij}}}],
\end{equation}
where $\bm{\Omega}_k\in \mathbb{R}^{h\times w}$, and $a_{k,ij}$ is the element in the $i$-th row and the $j$-th column of $\mathbf{A}_k$.
Then, the gradient matrix is operated by global average pooling, $\mathcal{A}(\cdot)$, and the obtained value is defined as the importance of $\mathbf{A}_k$, given as
\begin{equation} \label{GradMatrix}
	\omega_k=\mathcal{A}(\bm{\Omega}_k)=\frac{1}{hw} \sum_{i=1}^{h} \sum_{j=1}^{w} \frac{\partial \mathcal{L}(\mathbf{s}, \hat{\mathbf{s}})}{\partial a_{k,ij}}.
\end{equation}
Finally, the importance distribution vector of the feature tensor $\mathbf{A}$ can be represented as
\begin{equation}
	\bm{\omega}=[\omega_1, \omega_2, ..., \omega_c]^T.
\end{equation}
The detailed procedure is summarized in Algorithm \ref{FI}.

\begin{algorithm}[t]
	\begin{normalsize}
		\caption{Computing feature importance} 
		\label{FI}
		\DontPrintSemicolon
		\KwIn{The feature tensor $\mathbf{A}$, input image $\mathbf{s}$, and reconstructed image $\hat{\mathbf{s}}$.}
		\KwOut{The feature importance vector $\bm{\omega}$.}
		Calculate the system loss
		$\mathcal{L}(\mathbf{s}, \hat{\mathbf{s}})$ based on (\ref{Loss}).\\
		\For{$k\leftarrow 1$ \KwTo $c$}{
			Compute the gradient matrix $\bm{\Omega}_k$ based on (\ref{GradMatrix}).\\
			Apply average pooling to $\bm{\Omega}_k$ and obtain $\omega_k$.
		}
		Apply min-max normalization to $\bm{\omega}$.
	\end{normalsize}
\end{algorithm}

\subsubsection{Knowledge Distillation}
In practice, the application of Algorithm \ref{FI} faces challenges during the deployment phase.
This is primarily because both $\mathbf{A}$ and $\mathcal{L}(\mathbf{s}, \hat{\mathbf{s}})$ are essential inputs for the algorithm.
However, in communication systems, the system loss is typically unobservable until after transmission, rendering it infeasible to execute Algorithm \ref{FI} at the transmitter.
To tackle this issue, we aim to compute feature importance solely based on the features themselves, without prior knowledge of the system loss.
To achieve this, we devise an importance evaluator that leverages the knowledge distillation technique.

The knowledge distillation technique is a well-established method for transferring the insights and capabilities of a complex network, often referred to as the teacher model, into a more lightweight counterpart, known as the student model \cite{Hinton2015distilling}.
Drawing inspiration from this concept, we have successfully adapted the technique to transfer the knowledge required for calculating feature importance from the system model to a student model, which in our case is the importance evaluator.
This adaptation enables the importance evaluator to be effectively deployed during the operational phase.

\begin{figure}[t]
	\centering
	\includegraphics[width=0.47\textwidth]{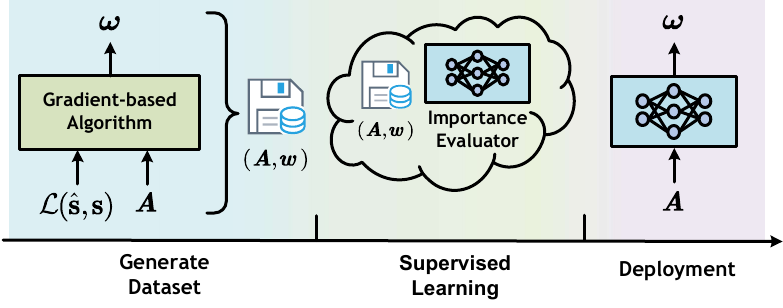}
	\captionsetup{font=footnotesize}
	\caption{Three stages of knowledge distillation.}
	\label{KD}
\end{figure}

As illustrated in Fig. \ref{KD}, the distillation process can be decomposed into three stages.
\begin{itemize}
	\item[(i)] Calculate feature importance using Algorithm \ref{FI}, and subsequently create a new dataset that pairs the resulting importance vector $\bm{\omega}$ with the corresponding feature tensor $\mathbf{A}$.
	\item[(ii)] Train the importance evaluator using the generated datasets via supervised learning.
	\item[(iii)] Replace Algorithm \ref{FI} with the importance evaluator during the deployment phase.
	
\end{itemize}

\subsection{Channel Predictor}
To facilitate feature allocation, prediction of future CSI becomes necessary.
Hence, the channel predictor estimates CSI samples and generates predictions accordingly.

To emulate correlated wide-sense stationary (WSS) Rayleigh fading channels, we utilize the enhanced sum-of-sinusoids (SOS) model described in \cite{Pop2001limitations}. Specifically, the $n$-th CSI sample, denoted as $\text{h}_n$, is expressed as
\begin{equation} \label{SOS}
	\text{h}_n = \text{h}(n T_s)=\frac{1}{\sqrt{M}} \sum_{m=1}^M[\text{I}_m(n T_s)+\mathrm{j} \text{Q}_m(n T_s)],
\end{equation}
where $T_s$ is the sampling period, $M$ is the number of multipaths, $\text{I}_m(n T_s)$ and $\text{Q}_m(n T_s)$ are the $m$-th in-phase and quadrature components, respectively, which are given as
\begin{align}  \label{IQComponents}
	\text{I}_m(n T_s) & = a_m \cos [(2 \pi f_d n T_s+\psi_m) \cos (\alpha_m)+\phi_m], \\
	\text{Q}_m(n T_s) & = b_m \sin [(2 \pi f_d n T_s+\psi_m) \cos (\alpha_m)+\phi_m],
\end{align}
where $a_m$ and $b_m$ are distributed as $\mathcal{N}(0, 1)$, $\alpha_m$ and $\phi_m$ denote the arrival angle and phase shift of the $m$-th path, respectively, and $f_d$ is the maximum Doppler shift in Hz.

Long short-term memory (LSTM) models belong to a category of DL networks featuring feedback loops, which enable them to preserve historical information \cite{Yu2019a}.
Their ability to maintain an internal state makes them exceptionally effective in tasks involving time series prediction.
Building on this capability, prior research has explored the integration of LSTM in channel prediction tasks, yielding significant achievements \cite{Peng2020lstm, Kibugi2021machine, Madhubabu2019}.
Consequently, we adopt the LSTM-based approach introduced in \cite{Madhubabu2019} to implement our channel predictor.

\subsection{Time Domain Feature Allocator}
To address the challenges posed by channel fluctuations in the time domain, we design a time domain feature allocator that utilizes the calculated feature importance $\bm{\omega}$ and the predicted CSI sequence $\tilde{\mathbf{h}}$.

\begin{figure}[t]
	\centering
	\includegraphics[width=0.48\textwidth]{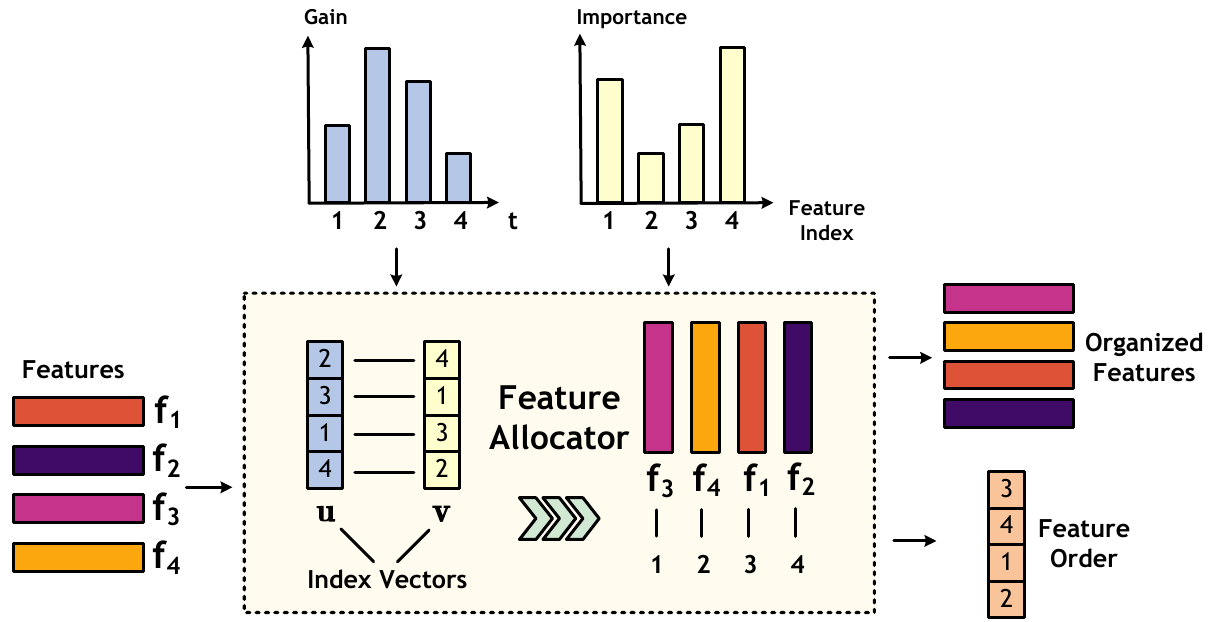}
	\captionsetup{font=footnotesize}
	\caption{The feature allocator.}
	\label{FeaArrMod}
\end{figure}

\begin{algorithm}[htb]
	\begin{normalsize}
		\caption{Time domain feature allocation} 
		\label{TFA}
		\DontPrintSemicolon
		\KwIn{The feature tensor $\mathbf{A}\in \mathbb{R}^{c\times h\times w}$, feature importance vector $\bm{\omega}\in \mathbb{R}^c$, and predicted CSI sequence $\tilde{\mathbf{h}} \in \mathbb{C}^c$.}
		\KwOut{The organized feature tensor $\tilde{\mathbf{A}}\in \mathbb{R}^{c\times h\times w}$ and feature order $\bm{\eta}\in \mathbb{R}^c$.}
		Mark each element in $\tilde{\mathbf{h}}$ and $\bm{\omega}$ with its original index.\\
		Compute the amplitude of each element in $\tilde{\mathbf{h}}$, i.e., $\tilde{\mathbf{h}}=abs(\tilde{\mathbf{h}})\in \mathbb{R}^c$.\\
		Sort $\tilde{\mathbf{h}}$ in descending order and obtain an index vector, $\mathbf{u}\in \mathbb{R}^c$.\\
		Sort $\bm{\omega}$ in descending order and obtain an index vector, $\mathbf{v}\in \mathbb{R}^c$.\\
		\For{$i\leftarrow 1$ \KwTo $c$}{
			$\,\bm{\eta}[\mathbf{u}[i]] = \mathbf{v}[i]$.\\
			$\tilde{\mathbf{A}}[\mathbf{u}[i]] = \mathbf{A}[\mathbf{v}[i]]$.
		}
	\end{normalsize}
\end{algorithm}

As exemplified in Fig. \ref{FeaArrMod}, the feature allocator follows a specific sequence of operations.
Initially, it marks each element in the feature importance vector $\bm{\omega}$ and the predicted CSI sequence $\tilde{\mathbf{h}}$ with the element's original index.
Then, it performs a descending sort on the elements in $\bm{\omega}$ and $\tilde{\mathbf{h}}$, and acquires two index vectors.
The larger the value of an element, the higher its index ranks in the index vector.
Subsequently, the allocator takes the index vectors and matches their elements with the same rank in pairs.
This procedure allocates each feature to its appropriate transmission time slot and establishes a new order in which the features are organized.
The newly determined feature order is then applied to the original feature tensor $\mathbf{A}$.
Finally, the allocator produces the organized feature tensor $\tilde{\mathbf{A}}$ along with the new feature order $\bm{\eta}$.
This feature order is also transmitted as supplementary information, incurring only a minimal increase in data overhead.
It is utilized by the inverse feature allocator at the receiver to reconstruct the initial feature tensor with the default feature order.
The detailed steps of the feature allocation algorithm are summarized in Algorithm \ref{TFA}.

\section{Space-Time Feature Allocation} \label{SecSpaceTimeAlloc}
In this section, we provide a more detailed explanation of space-time domain feature allocation.
We focus on a scenario with $N_t$ transmit and $N_r$ receive antennas.
Unlike the single-antenna case, the CSI matrix is denoted as $\mathbf{H}\in \mathbb{C}^{N_r\times N_t}$.

Cross-transmission commonly occurs in MIMO channels, resulting in considerable interference among antennas.
The semantic features transmitted via various transmit antennas experience varying degrees of inter-antenna interference and channel fluctuations in the spatial domain.
In traditional communication, two types of MIMO systems are prevalent: those that operate without precoding and those that rely on precoding techniques.
However, within the framework of semantic communication, these conventional approaches cannot provide optimal performance, as they do not account for semantic importance.
Hence, we introduce space-time feature allocation as a means to significantly enhance their performance.

\subsection{Feature Allocation for Precoding-Free Systems}
We begin by examining a MIMO semantic communication system that operates without employing precoding techniques.
\subsubsection{Equalization at the Receiver}
MMSE equalization is a common MIMO detection technique, offering a means to mitigate channel interference via post-processing at the receiver \cite{Yang2015fifty}.
An MMSE equalizer is utilized at the receiver's end to estimate the transmitted signal.
This equalizer is designed to minimize the MSE between the transmitted and detected signals, expressed as
\begin{equation}
	\mathbf{G}^*=\mathop{\arg\min}\limits_{\mathbf{G}} \mathbb{E}_{\mathbf{x}}\{||\mathbf{x}-\mathbf{Gy}||^2\}.
\end{equation}
The optimal matched filter can be derived as
\begin{equation}\label{Equalizer}
	\mathbf{G}^*=\mathbf{H}^H(\mathbf{HH}^H+\frac{\sigma^2}{P}\mathbf{I})^{-1},
\end{equation}
where $\sigma^2$ and $P$ are the variance of the AWGN and the power of the transmitted signal, respectively.
Unless otherwise specified, we will use $\mathbf{G}$ to denote the optimal matched filter in the following text.
\subsubsection{Transmission Procedure}
We will now provide further details on the adjusted transmission process within the context of the FAST framework outlined in Section \ref{SecFramework}.
Initially, each feature is sent via a dedicated transmit antenna during a specific time slot.
As a result, the channel input symbols $\mathbf{x}$ are partitioned into sets of signals, with each set comprising $N_t$ symbols.
Taking the $i$-th set $\mathbf{x}_i\in \mathbb{C}^{N_t}$ as an example, the received signal is represented as
\begin{equation}
	\mathbf{y}_i=\mathbf{Hx}_i+\mathbf{n}.
\end{equation}
The receiver estimates the CSI matrix $\mathbf{H}$, which is leveraged to compute the detector $\mathbf{G}$ based on (\ref{Equalizer}).
Thus, the detected signal is given by
\begin{equation}
	\hat{\mathbf{x}}_i=\mathbf{Gy}_i=\mathbf{GHx}_i+\mathbf{Gn}.
\end{equation}
\subsubsection{Feature Allocation over Transmit Antennas}
In practice, the features transmitted by different transmit antennas exhibit variations in the received SINR, which serves as an indicator of the transmit antenna's performance.
To illustrate this, consider the $j$-th transmit antenna, denoted as $\text{TX}_j$.
Its received SINR can be calculated as
\begin{align}\label{SINR}
	\text{SINR}(\text{TX}_j) & =\frac{||\mathbf{g}_j\mathbf{h}_j\text{x}_{ij}||^2}{\sum\limits_{k\neq j}||\mathbf{g}_j\mathbf{h}_k\text{x}_{ij}||^2+||\mathbf{g}_j\mathbf{n}||^2} \notag \\
	& = \frac{P||\mathbf{g}_j\mathbf{h}_j||^2}{P\sum\limits_{k\neq j}||\mathbf{g}_j\mathbf{h}_k||^2+||\mathbf{g}_j||^2\sigma^2}.
\end{align}
Here,  $\mathbf{g}_j\in \mathbb{C}^{1\times N_r}$ represents the $j$-th row vector of $\mathbf{G}$, $\mathbf{h}_j\in \mathbb{C}^{N_r\times 1}$ represents the $j$-th column vector of $\mathbf{H}$, and $\text{x}_{ij}$ signifies the symbol transmitted from the $i$-th set by TX$_j$.
According to (\ref{Equalizer}) and (\ref{SINR}), the received SINR of $\text{TX}_j$ is related to $j$, $\mathbf{H}$, $P$, and $\sigma^2$.
Note that $P$ and $\sigma^2$ are both constants, thus the received SINR fluctuates with variations in the transmit antenna and the CSI.
This uncertainty can cause a crucial feature to experience a low received SINR, consequently impairing the overall system performance.

To cope with the uncertainty, we first introduce the concept of a space-time resource block, denoted as a particular transmit antenna within a specific time slot, whose quality is determined by the associated received SINR.
Then, an extended space-time feature allocator is developed.
Since we consider the space-time domain, the predicted CSI sequence becomes $\tilde{\mathbf{H}}=[\tilde{\mathbf{H}}_{t+1}, ..., \tilde{\mathbf{H}}_{t+\frac{c}{N_t}}]$.
It is exploited to compute the corresponding equalizer sequence, represented as $\tilde{\mathbf{G}}=[\tilde{\mathbf{G}}_{t+1}, ..., \tilde{\mathbf{G}}_{t+\frac{c}{N_t}}]$.
Subsequently, a block quality matrix, signified as $\tilde{\mathbf{Q}}\in\mathbb{R}^{N_t\times \frac{c}{N_t}}$, is computed at the receiver on the basis of (\ref{SINR}),
given as
\begin{equation}
	\tilde{\mathbf{Q}}=\begin{bmatrix}
		\tilde{\text{q}}_{1, t+1} & \tilde{\text{q}}_{1, t+2} & \cdots & \tilde{\text{q}}_{1, t+\frac{c}{N_t}}\\
		\tilde{\text{q}}_{2, t+1} & \tilde{\text{q}}_{2, t+2} & \cdots & \tilde{\text{q}}_{2, t+\frac{c}{N_t}}\\
		\vdots & \vdots & \ddots & \vdots \\
		\tilde{\text{q}}_{N_t, t+1} & \tilde{\text{q}}_{N_t, t+2} & \cdots & \tilde{\text{q}}_{N_t, t+\frac{c}{N_t}}\\
	\end{bmatrix}_{N_t\times \frac{c}{N_t}},
\end{equation}
where $\text{q}_{ij}(i=1,2,...,N_t;j=1,2,...,\frac{c}{N_t})$ denotes the received SINR for $\text{TX}_i$ in the $j$-th time slot.
The matrix $\tilde{\mathbf{Q}}$ is then flattened into a vector, $\tilde{\mathbf{q}}\in \mathbb{R}^c$.
Similar to the time domain designs, the receiver sorts $\tilde{\mathbf{q}}$ in descending order and obtains an index vector, $\mathbf{u}$.
Finally, only the index vector $\mathbf{u}$ is transmitted back to the transmitter via the feedback channel, requiring only a small number of bits.
The details of the allocation procedure are summarized in Algorithm \ref{STFANprecoding}.
\begin{algorithm}[t]
	\begin{normalsize}
		\caption{Space-time domain feature allocation algorithm for systems without precoding} 
		\label{STFANprecoding}
		\DontPrintSemicolon
		\KwIn{The feature tensor $\mathbf{A}\in \mathbb{R}^{c\times h\times w}$, feature importance vector $\bm{\omega}\in \mathbb{R}^c$, predicted CSI sequence $\tilde{\mathbf{H}} \in \mathbb{C}^{\frac{c}{N_t}\times N_r\times N_t}$, transmitting power $P$, and AWGN variance $\sigma^2$.}
		\KwOut{The organized feature tensor $\tilde{\mathbf{A}}\in \mathbb{R}^{c\times h\times w}$ and feature order $\bm{\eta}\in \mathbb{R}^c$.}
		Compute $\tilde{\mathbf{G}}$ based on $\tilde{\mathbf{H}}$, $P$, and $\sigma^2$, using (\ref{Equalizer}).\\
		Compute $\tilde{\mathbf{Q}}$ based on $\tilde{\mathbf{G}}$, $\tilde{\mathbf{H}}$, $P$, and $\sigma^2$, using (\ref{SINR}).\\
		Flatten $\tilde{\mathbf{Q}}$ into a vector, $\tilde{\mathbf{q}}$.\\
		Mark each element in $\tilde{\mathbf{q}}$ and $\bm{\omega}$ with its original index.\\
		Sort $\tilde{\mathbf{q}}$ in descending order to obtain the index vector, $\mathbf{u}\in \mathbb{R}^c$.\\
		The receiver transmits $\mathbf{u}$ back to the transmitter.\\
		The transmitter sorts $\bm{\omega}$ in descending order to obtain the index vector $\mathbf{v}\in \mathbb{R}^c$.\\
		\For{$i\leftarrow 1$ \KwTo $c$}{
			$\,\bm{\eta}[\mathbf{u}[i]] = \mathbf{v}[i]$.\\
			$\tilde{\mathbf{A}}[\mathbf{u}[i]] = \mathbf{A}[\mathbf{v}[i]]$.
		}
	\end{normalsize}
\end{algorithm}
\subsection{Feature Allocation for Precoding-Based Systems}
Next, we examine a system based on precoding techniques.
\subsubsection{Precoding at the Transmitter}
In conventional MIMO communication systems, precoding plays a pivotal role leveraging knowledge of the CSI to address channel impairments by manipulating the signal before transmission \cite{Vu2007mimo}.
Earlier studies have established the effectiveness of integrating the classical SVD precoding technique into MIMO semantic communication systems \cite{O2017deep, Zhang2022svd, Zhang2023scan}.
Building upon these insights, we integrate SVD precoding into the framework of space-time feature allocation.

\begin{figure}[t]
	\centering
	\includegraphics[width=0.35\textwidth]{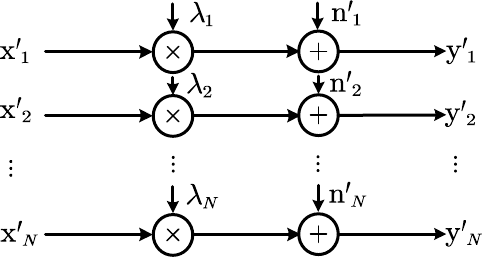}
	\captionsetup{font=footnotesize}
	\caption{The equivalent parallel complex Gaussian channel.}
	\label{EqChannel}
\end{figure}

As shown in Fig. \ref{EqChannel}, through SVD precoding, the MIMO channel can be simplified into an equivalent set of parallel subchannels.
Each of these subchannels can be regarded as a scalar Gaussian channel with gain $\lambda_i\in \mathbb{R}$ and AWGN $\text{n}'_i\in \mathbb{C}$ $(i=1,2,...,N;N=\text{min}(N_r, N_t))$.
Mathematically, this can be expressed as
\begin{equation}
	\mathbf{H}=\mathbf{UDV}^H,
\end{equation}
where $\mathbf{U}\in\mathbb{C}^{N_r\times N_r}$ and $\mathbf{V}^H\in\mathbb{C}^{N_t\times N_t}$ are unitary matrices, and $\mathbf{D}$ is a diagonal matrix containing the singular values of $\mathbf{H}$ along its main diagonal.
Specifically, we have 
\begin{equation}
	\mathbf{D}=\begin{bmatrix}
		\mathbf{S} & \mathbf{O} \\
		\mathbf{O} & \mathbf{O} \\
	\end{bmatrix}_{N_r\times N_t},
\end{equation}
where $\mathbf{S}=\text{diag}(\lambda_1, \lambda_2, ..., \lambda_N)$ and $\mathbf{O}$ denotes a zero matrix.
\subsubsection{Transmission Procedure}
Next, we will delineate the alterations to the transmission procedure as discussed in Section \ref{SecFramework}.
Initially, each feature is sent through an individual parallel subchannel.
Consequently, the channel input symbols are divided into sets of signals, with each set containing $d$ symbols, where $d$ represents the number of data streams.
Considering the $i$-th set $\mathbf{x}_i \in\mathbb{C}^d$, it is then multiplied by SVD-based precoder $\mathbf{V}\in \mathbb{C}^{N_r\times d}$.
Afterwards, the power constraint $P$ is imposed to the precoded symbols, i.e., $||\mathbf{Vx}_i||^2\leq P$.
Then, the received signal is given by
\begin{equation}
	\mathbf{y}_i=\mathbf{HVx}_i+\mathbf{n}.
\end{equation}
Finally, the transmitted symbols are detected at the receiver by
\begin{equation}
	\hat{\mathbf{x}}_i=\mathbf{U}^H\mathbf{y}_i=\mathbf{U}^H\mathbf{HVx}_i+\mathbf{U}^H\mathbf{n}.
\end{equation}
\subsubsection{Feature Allocation over Subchannels}
The singular values of the channel matrix $\mathbf{H}$ demonstrate disparities in the gains of the subchannels.
Consequently, despite the ability of precoding techniques to alleviate channel impairments, it is crucial to recognize that essential features may be received with low quality through the weak subchannels.
This can, in turn, result in a deterioration of the overall system performance.

Fortunately, building upon the concept of feature allocation in the time domain, we can effectively address this challenge by assigning each feature to the most suitable subchannel.
In precoding-based systems, a space-time resource block is identified as a specific subchannel within a particular time slot.
The quality of this resource block is determined by its associated singular value.
To address this, we have designed a space-time feature allocator customized for precoding-based systems.
Initially, the allocator employs the SVD algorithm on the predicted CSI sequence, $\tilde{\mathbf{H}}=[\tilde{\mathbf{H}}_{t+1}, ..., \tilde{\mathbf{H}}_{t+\frac{c}{N}}]$, and obtains a sequence of diagonal matrices, $\tilde{\mathbf{D}}=[\tilde{\mathbf{D}}_{t+1}, ..., \tilde{\mathbf{D}}_{t+\frac{c}{N}}]$.
Afterwards, the block quality matrix $\tilde{\mathbf{Q}}$ is constructed by stacking together the singular values from the diagonal matrices, given by
\begin{equation}
	\tilde{\mathbf{Q}}=\begin{bmatrix}
		\tilde{\lambda}_{1, t+1} & \tilde{\lambda}_{1, t+2} & \cdots & \tilde{\lambda}_{1, t+\frac{c}{N}}\\
		\tilde{\lambda}_{2, t+1} & \tilde{\lambda}_{2, t+2} & \cdots & \tilde{\lambda}_{2, t+\frac{c}{N}}\\
		\vdots & \vdots & \ddots & \vdots \\
		\tilde{\lambda}_{N, t+1} & \tilde{\lambda}_{N, t+2} & \cdots & \tilde{\lambda}_{N, t+\frac{c}{N}}\\
	\end{bmatrix}_{N\times \frac{c}{N}}.
\end{equation}
The matrix $\tilde{\mathbf{Q}}$ is then flattened into a vector and sorted in descending order along with the importance vector $\bm{\omega}$.
Each element is also marked with its original index.
The detailed procedure is similar to the space-time feature allocation for precoding-free systems, as elaborated in Algorithm \ref{STFANprecoding}.

\section{Simulation Results} \label{Simulation}
The simulations are conducted using the PyTorch platform.
We employ the convolutional neural network (CNN)-based autoencoder proposed in \cite{Bourtsoulatze2019deep} as the architecture of the basic model.
In addition, the framework of the importance evaluator is also based on CNN, while the channel predictor is constructed with LSTM layers.
The detailed settings of the employed networks are presented in Table \ref{Setting}.
For time domain feature allocation, we consider a single-antenna case, where the improved SOS model in \cite{Pop2001limitations} is employed as the physical channel.
For space-time domain feature allocation, we consider two multiple-antenna cases with $N_t=2$ transmit antennas and $N_r=2$ receive antennas, while the MIMO channel is constructed by the combination of $N_r\times N_t$ links generated by the SOS model.
For the precoding-based system, the number of data streams is set as $d=2$.
To train the basic models, the ``Adam'' optimizer is utilized with an initial learning rate of 0.001.
The learning rate is gradually reduced by the cosine annealing algorithm \cite{Loshchilov2016sgdr} as training progresses.
The batch size is set to 64.
We start by evaluating our FAST approach on the CIFAR-10 dataset, which consists of $50,000$ RGB images of size $32\times 32$ in the training dataset and $10,000$ in the testing dataset.
For higher resolution images, we adopt the ImageNet dataset for training and perform evaluation on the Kodak dataset.
ImageNet consists of $1.2$ million images of random size, which are randomly cropped into patches of size $96\times 96$ during model training.
Kodak is composed of $24$ images of size $768\times 512$.

Following \cite{Bourtsoulatze2019deep}, for the single-antenna case, we define the image size $l$, the channel input size $k$, and $R = k/l$ as the source bandwidth, the channel bandwidth, and the bandwidth ratio, respectively.
For the multiple-antenna cases, we redefine the bandwidth ratio as $R = k/(lN_t)$ and $R = k/(ld)$ for the precoding-free and precoding-based systems, respectively.
The SNR is defined as $10 \log _{10} (P/\sigma^2)(\text{dB})$, where $P$ is the transmit power and $\sigma^2$ is the noise variance at the receiver.
In the single-antenna case, the basic model is trained at a bandwidth ratio of $R = 1/4$, and three separate models are developed and tailored to three different SNRs:  $\text{SNR}_\text{train}=$ $7$ dB, $13$ dB, $19$ dB.
In the multiple-antenna cases, the training bandwidth ratio $R$ is set to $1/8$, and performance is evaluated at values of $\text{SNR}_\text{test}$ from $0$ to $25$ dB.
To evaluate the performance of the system, we employ the peak signal-to-noise ratio (PSNR) metric, which is defined as
\begin{equation}
	\text{PSNR}=10 \log _{10} \frac{\text{MAX}^2}{\text{MSE}}(\text{dB}),
\end{equation}
where $\text{MSE}=\frac{1}{l}\|\mathbf{s}-\hat{\mathbf{s}}\|^2$ and $\text{MAX}$ is the maximum possible value of an image pixel.
In addition, we also use the structural similarity index (SSIM) metric \cite{Zhou2004image} in our evaluations, which can effectively capture the perceived visual quality of reconstructed images.

\renewcommand{\arraystretch}{1.5}
\begin{table}[t]\small
	\centering
	\caption{Settings of the employed networks.}  
	\label{Setting}
	\begin{tabular}{|c|c|c|}
		\hline
		& Layer Name & Dimension  \\
		\hline
		\multirow{3}{*}{\thead{Semantic \\ Encoder}}
		& ConvLayer & 16 (kernels) \\
		\cline{2-3}
		& $3\times$ ConvLayer & 32 (kernels) \\
		\cline{2-3}
		& ConvLayer & 24 (kernels)  \\
		
		\hline
		\multirow{3}*{\thead{Semantic \\ Decoder}}
		& $3\times$ TransConvLayer & 32 (kernels) \\
		\cline{2-3}
		& TransConvLayer & 16 (kernels) \\
		\cline{2-3}
		& TransConvLayer & 3 (kernels) \\
		
		\hline
		\multirow{2}*{\thead{Channel \\ Predictor}}
		& $3\times$LSTM Layer & 120  \\
		\cline{2-3}
		& Dense & 2 \\
		
		\hline
		\multirow{2}*{\thead{Importance \\ Evaluator}}
		& $3\times$ ConvLayer & 24 (kernels) \\
		\cline{2-3}
		& $2\times$ Dense & 24  \\
		\hline
	\end{tabular}
\end{table}

\begin{figure}[htb]
	\centering
	\subfloat[Single antenna case.]{\includegraphics[width=0.45\textwidth]{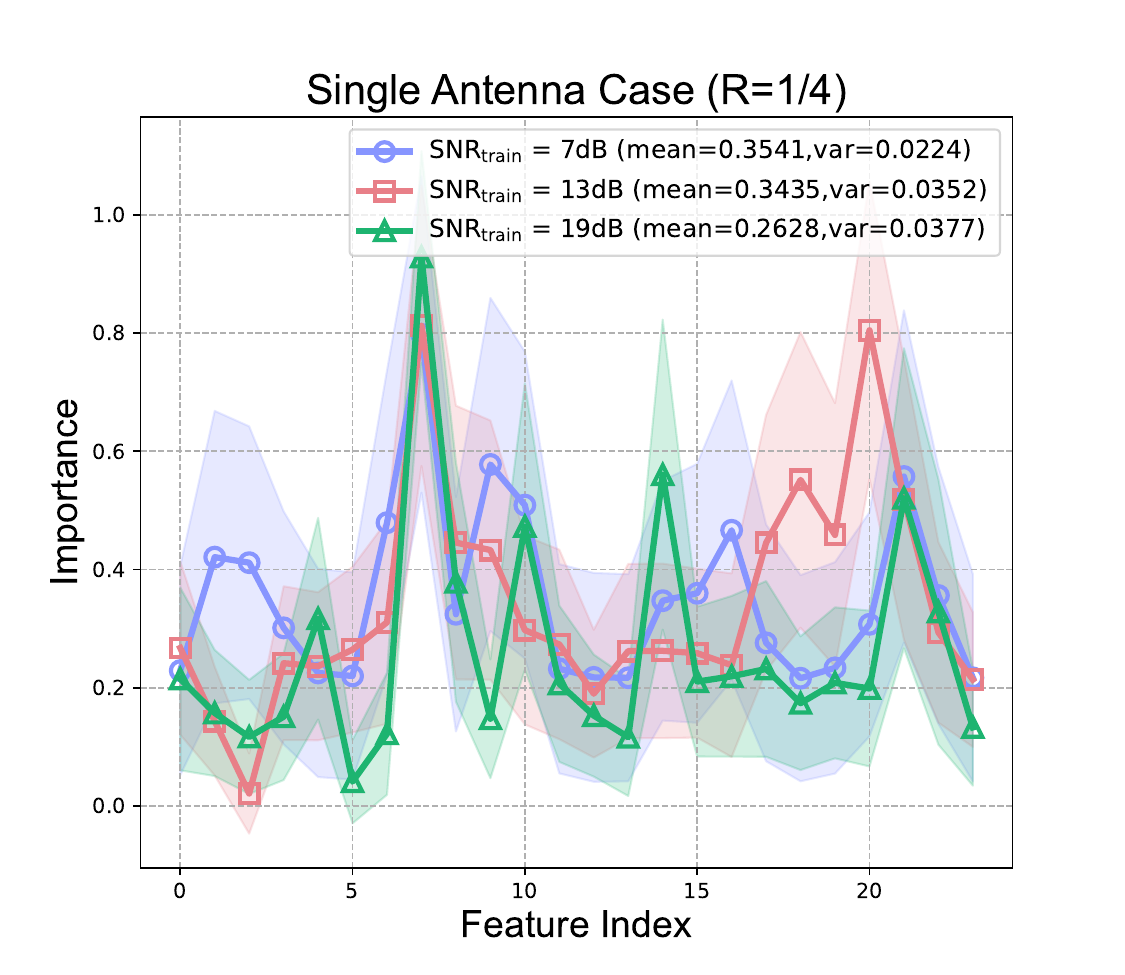}\label{TFI}}

	\subfloat[Multiple antenna case.]{\includegraphics[width=0.45\textwidth]{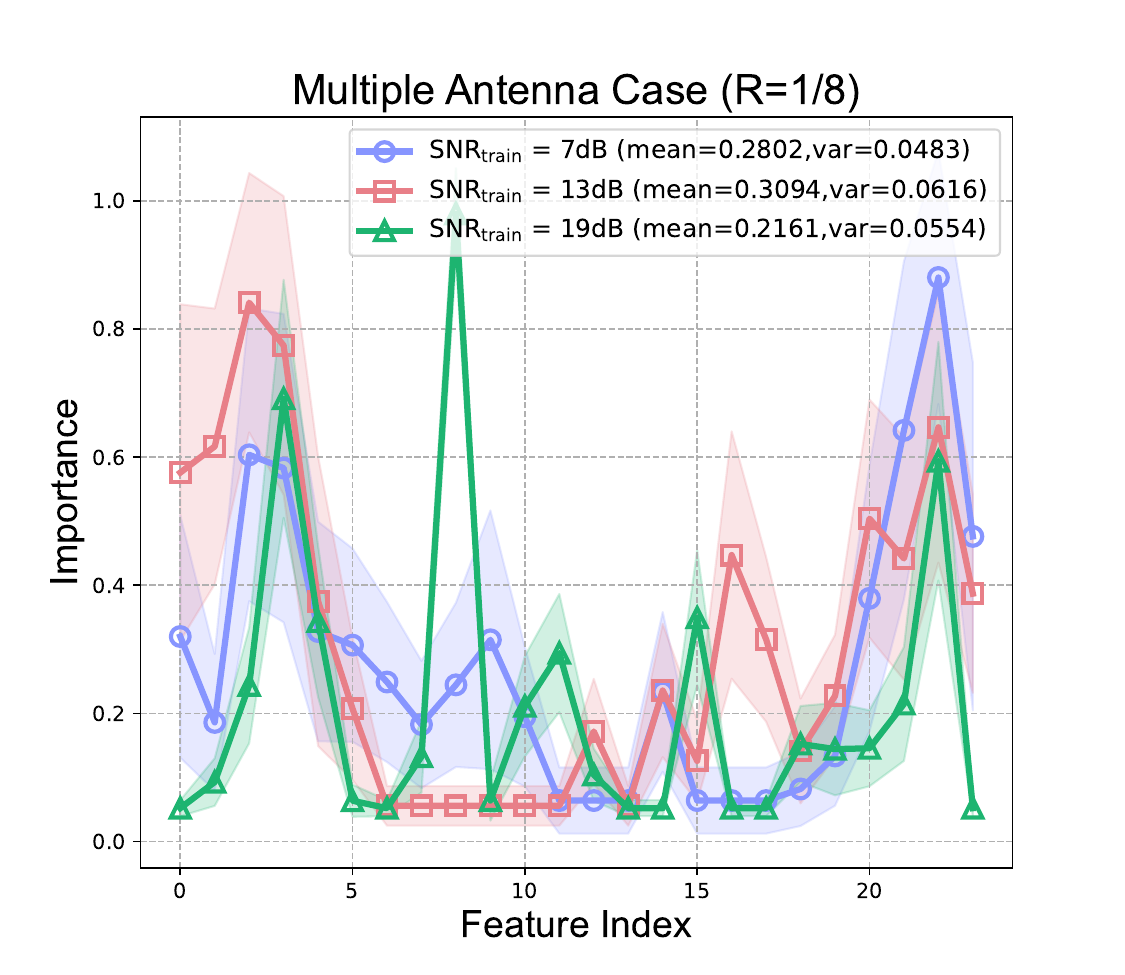}\label{STFI}}
	\captionsetup{font=footnotesize}
	\caption{The importance distribution of the features encoded by $3$ different basic models. Each data point on the curves signifies the mean value of feature importance across the entire CIFAR-10 dataset, while the transparent region surrounding a data point of the same color represents its standard deviation.}
	\label{DFI}
\end{figure}

\subsection{Distribution of Feature Importance}
All image samples in the CIFAR-10 dataset are encoded into semantic features using different pre-trained basic models.
In particular, with $k/l=1/4$, each image is encoded into $24$ features.
Afterwards, we calculate the importance of each feature using Algorithm \ref{FI}, and obtain the distribution of feature importance through statistical analysis.

Fig. \ref{DFI} presents the importance distribution of the features encoded by the $3$ basic models trained at different SNRs, where Fig. \ref{DFI}(a) illustrates the single-antenna case and Fig. \ref{DFI}(b) illustrates the multiple-antenna case.
Each data point on the curves signifies the mean of the feature importance across the entire dataset, while the transparent region surrounding a data point of the same color represents its standard deviation.
Some inferences can be derived as follows:
\begin{itemize}
	\item The results suggest that, as $\text{SNR}_\text{train}$ increases, the importance of features across different images tends to converge toward the mean values (i.e., the solid curve).
	This implies that feature importance becomes less sensitive to the source data at higher $\text{SNR}_\text{train}$.
	\item It is readily seen that the importance curve generally exhibits more pronounced fluctuations as $\text{SNR}_\text{train}$ increases, a trend that is also reflected in the variance annotations in the legend.
	This suggests that feature importance is more sensitive to higher $\text{SNR}_\text{train}$, aligning with our intuitive expectations.
	When the channel state is favorable, only a limited number of features are necessary for the task, while the remaining ones are redundant.
	This leads to a polarization of feature importance.
	As the channel state deteriorates, more features become required, resulting in a convergence of feature importance scores.
	\item Moreover, the importance curves associated with lower $\text{SNR}_\text{train}$ tend to have higher mean values.
	This is due to the fact that more information must be transmitted when the channel state is poor.
	As such, the model tends to have the features encoded more equally. 
	Consequently, the features generally exhibit higher importance, leading to higher mean values.
\end{itemize}

\subsection{Effectiveness of Importance Evaluator}
\begin{figure}[t]
	\centering
	\subfloat[$\text{R}=1/4,\text{SNR}_\text{train}=7$ dB]{\includegraphics[width=0.25\textwidth]{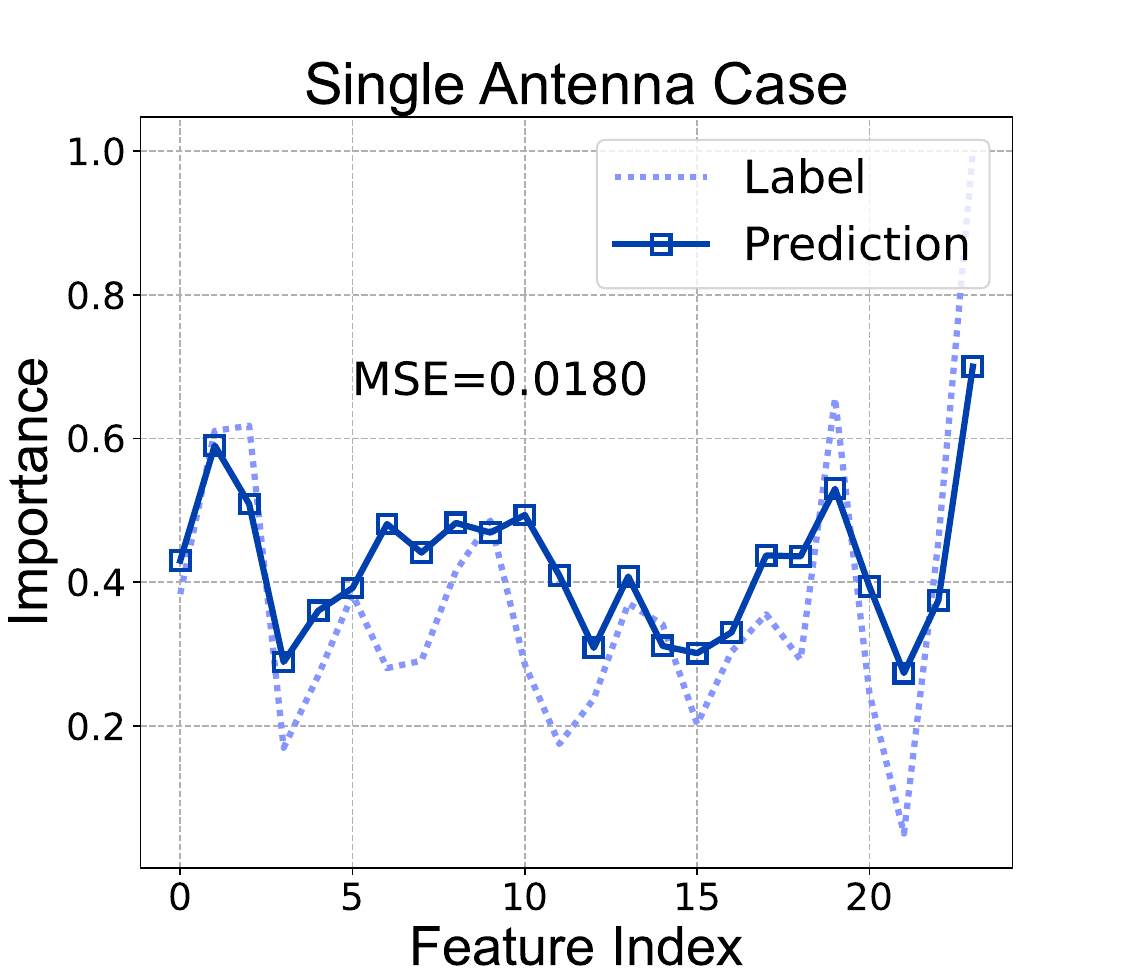}}
	\subfloat[$\text{R}=1/8,\text{SNR}_\text{train}=7$ dB]{\includegraphics[width=0.25\textwidth]{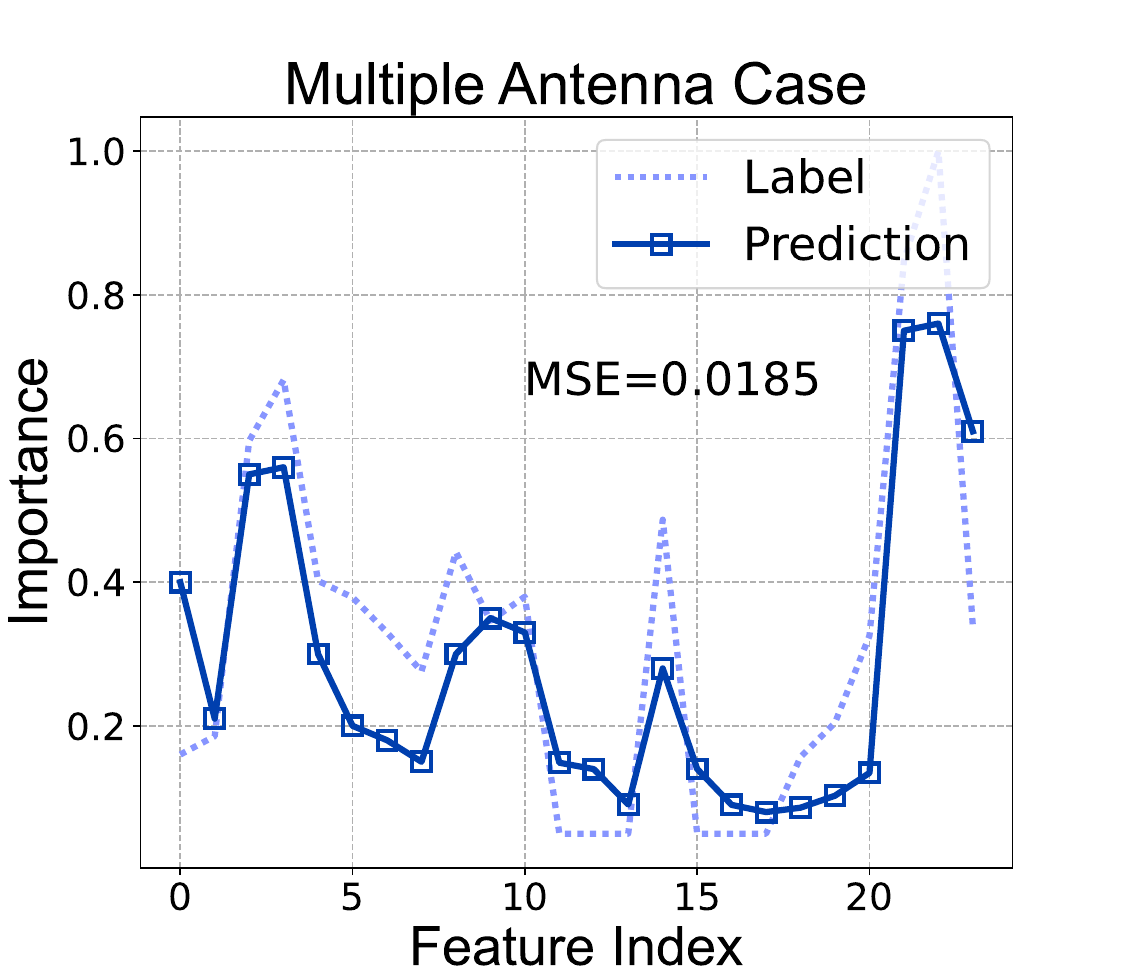}}
	\hfill
	\subfloat[$\text{R}=1/4,\text{SNR}_\text{train}=13$ dB]{\includegraphics[width=0.25\textwidth]{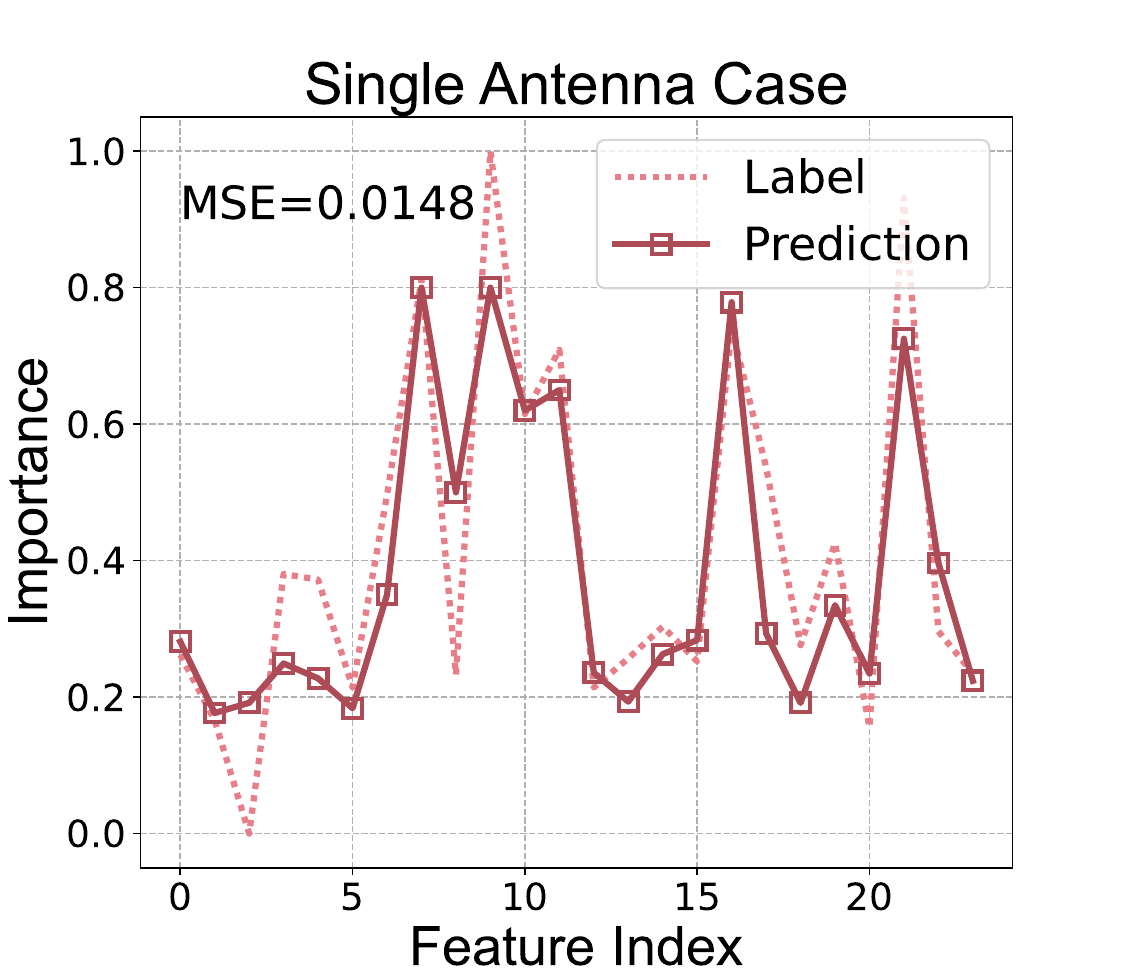}}
	\subfloat[$\text{R}=1/8,\text{SNR}_\text{train}=13$ dB]{\includegraphics[width=0.25\textwidth]{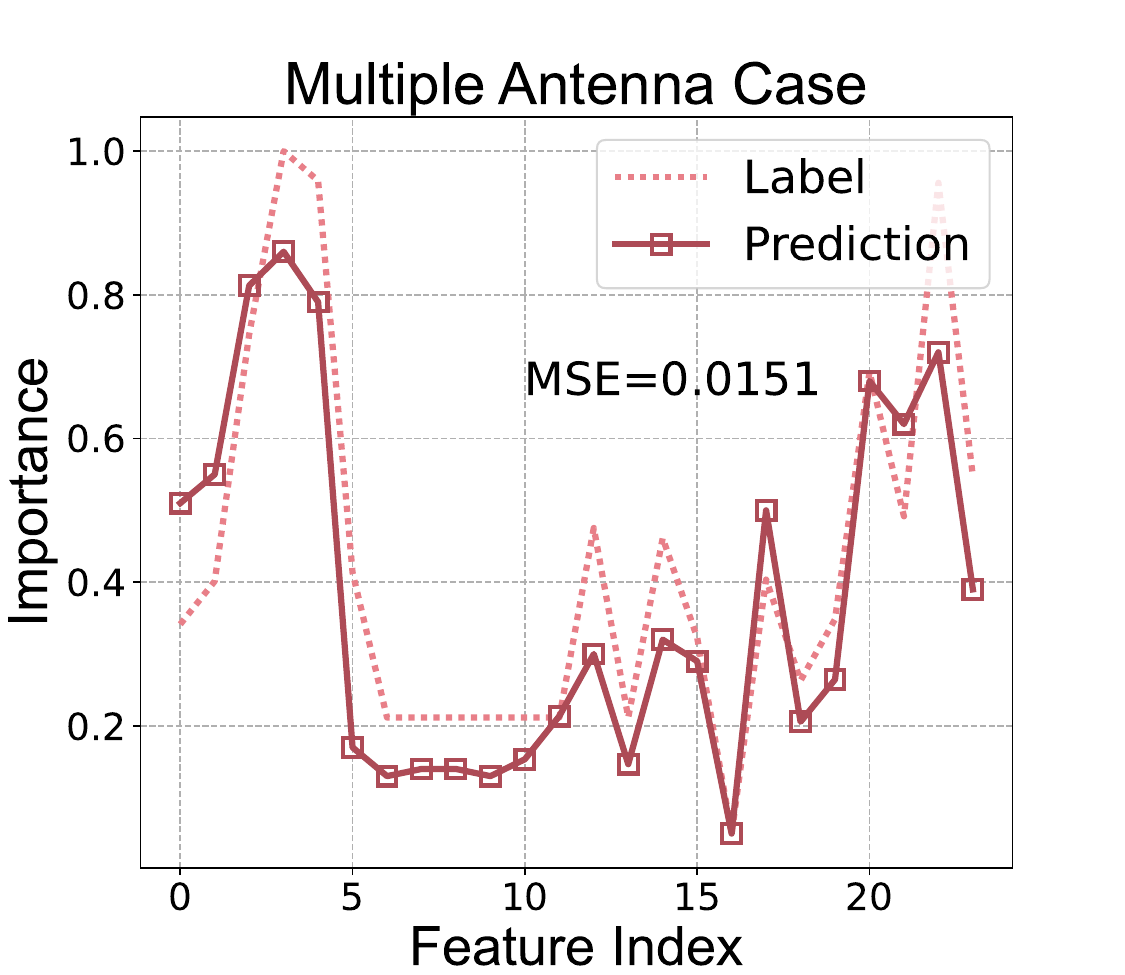}}
	\hfill
	\subfloat[$\text{R}=1/4,\text{SNR}_\text{train}=19$ dB]{\includegraphics[width=0.25\textwidth]{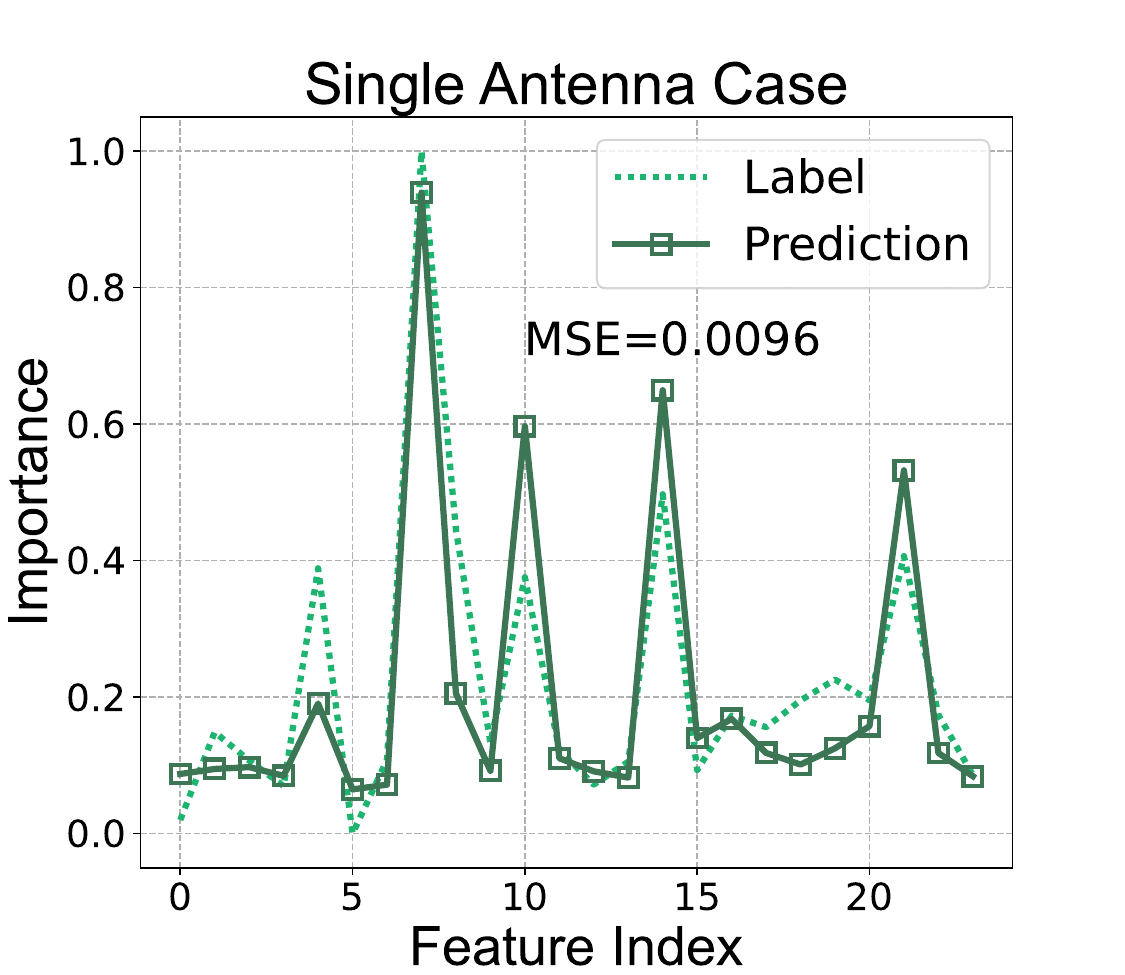}}
	\subfloat[$\text{R}=1/8,\text{SNR}_\text{train}=19$ dB]{\includegraphics[width=0.25\textwidth]{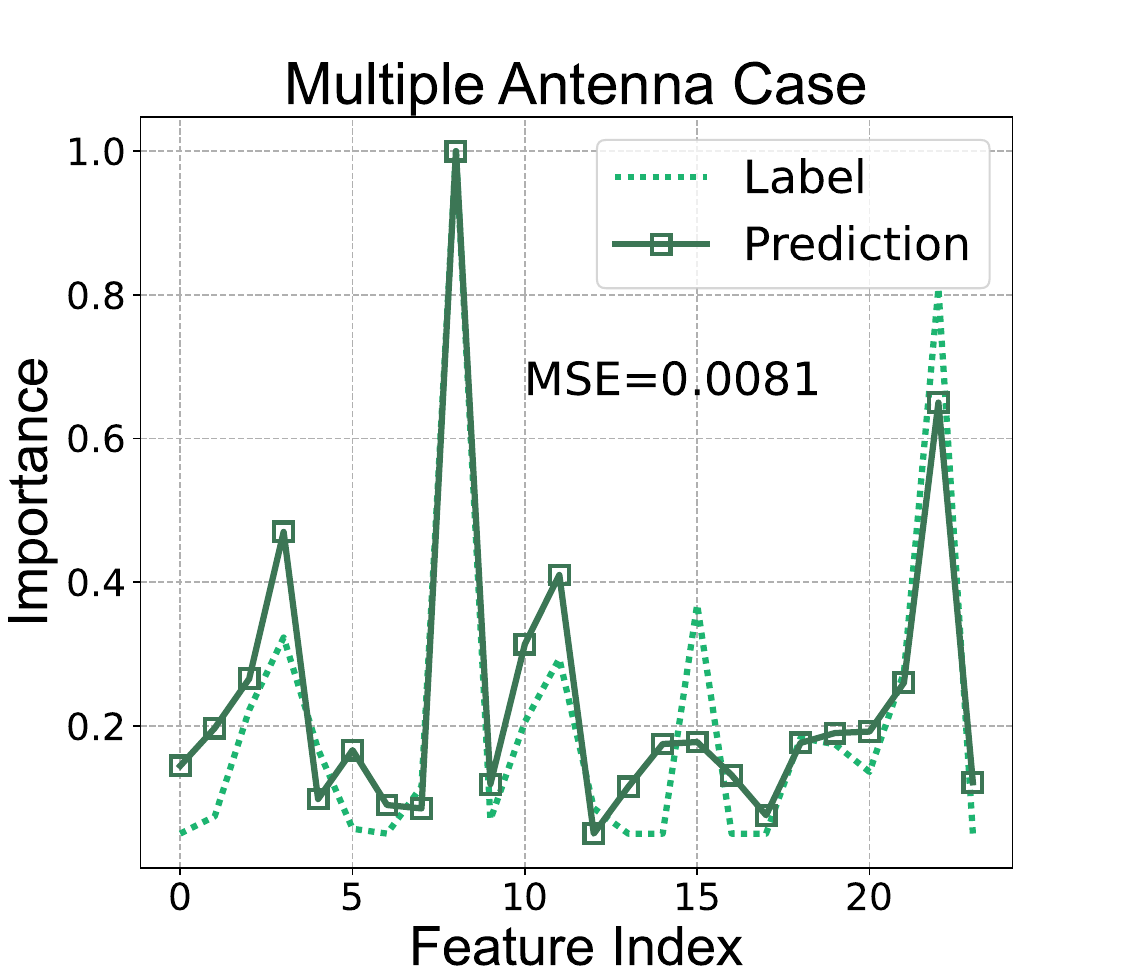}}
	\captionsetup{font=footnotesize}
	\caption{The performance of the importance evaluators tailored for different pre-trained basic models. The left column presents the single-antenna case and the right column presents the multiple-antenna case.}
	\label{PIE}
\end{figure}

For each basic model trained at a specific SNR, a corresponding importance evaluator is also trained.
To demonstrate the effectiveness of the importance evaluator, we begin by randomly selecting an image from the test dataset.
Then, we encode it into semantic features using different basic models pre-trained at different SNRs, and calculate the importance of the features using Algorithm \ref{FI}.
Subsequently, we feed the features to the importance evaluators that have distilled knowledge from the basic models, and obtain the predicted importance.
In the final step, we compare the predicted importance scores to the calculated ones, which are regarded as labels.
This comparison is then visualized to illustrate the effectiveness of the proposed importance evaluator.
Tt is important to highlight that for achieving feature allocation, our primary concern lies in the relative ranking of importance scores, rather than their exact numerical values.
Consequently, as long as the predicted scores are similar to the ordering of the label scores, the effectiveness of the importance evaluator is verified.

The results are shown in Fig. \ref{PIE}, where the left column presents the single-antenna case and the right column presents the multiple-antenna case.
It is evident that the proposed importance evaluator can provide accurate predictions for the feature importance scores.
The accuracy exhibits an upward trend as $\text{SNR}_\text{train}$ increases, which corresponds to the inferences drawn from Fig. \ref{DFI}.
Particularly, as $\text{SNR}_\text{train}$ increases, the importance scores become less sensitive to source data, thereby making it easier for the evaluator to perform prediction across different images.
Furthermore, we can also observe that the prediction curves exhibit a similar trend to the corresponding mean value curves depicted in Fig. \ref{DFI}.
These observations suggest that the importance evaluator can effectively learn the importance distribution of semantic features.

\subsection{Performance of FAST in the Time Domain}
We compare the performance of the following schemes:
\begin{itemize}
	\item JSCC+FAST(PC+IE): The proposed FAST approach, where the CSI is estimated via channel prediction (PC), and the feature importance is calculated by the importance evaluator (IE).
	\item JSCC+FAST(KC+IE): An ideal variant of FAST assuming precisely known future CSI (KC).
	\item JSCC+FAST(KC): A variant of FAST assuming precise CSI without employing the knowledge distillation technique.
	\item JSCC+FAST(PC): A variant of FAST without the knowledge distillation technique.
	\item JSCC: A benchmark system that only uses the pre-trained basic model, without employing feature allocation.
\end{itemize}

\begin{figure}[t]
	\centering
	\includegraphics[width=0.45\textwidth]{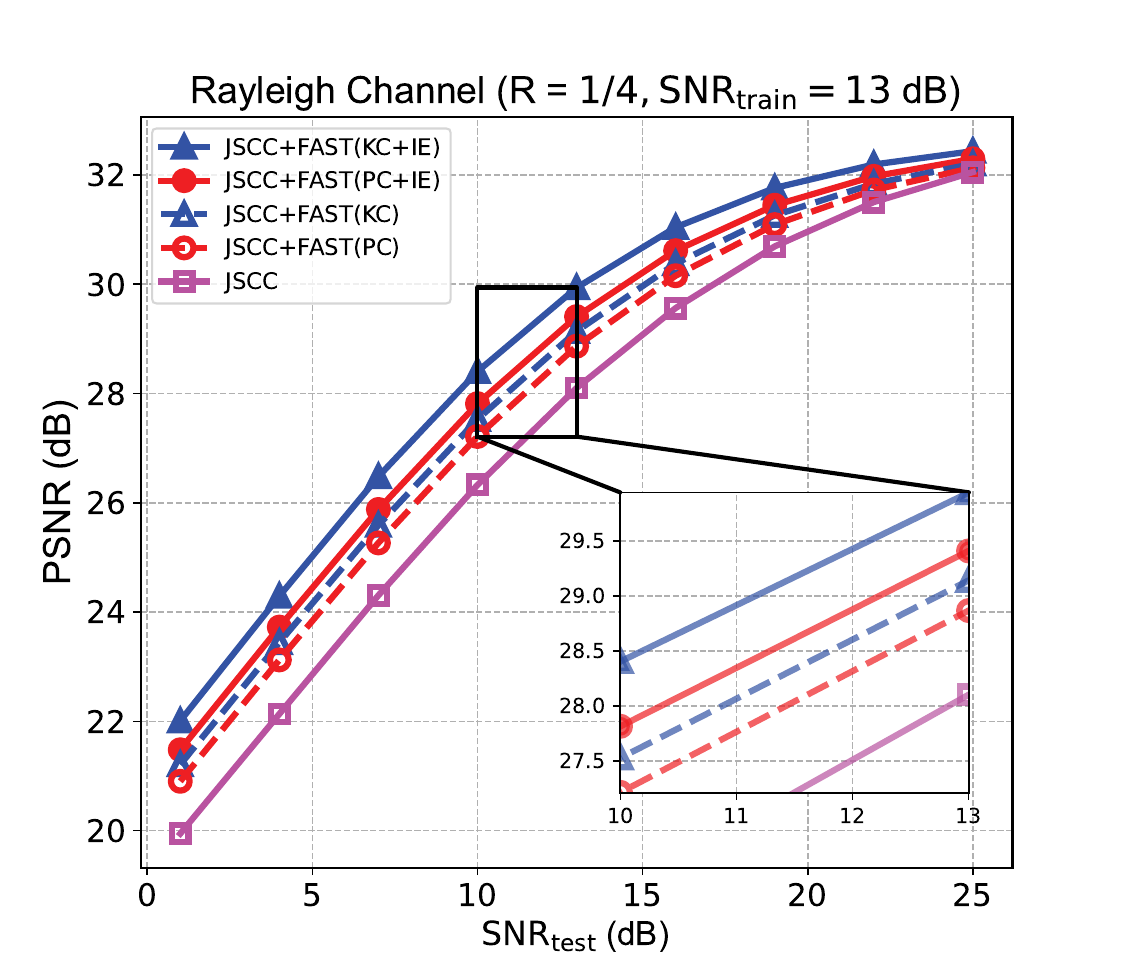}
	\captionsetup{font=footnotesize}
	\caption{Comparison of different schemes in the time domain. The basic models are all trained with a SISO Rayleigh channel at $\text{R} = 1/4$ and $\text{SNR}_{\text{train}} = 13$ dB. The performance of these schemes is evaluated on the CIFAR-10 dataset by calculating the average PSNR of $10,000$ images.}
	\label{Comparison}
\end{figure}

Fig. \ref{Comparison} shows the performance of different schemes versus $\text{SNR}_\text{test}$.
It is readily seen that FAST and all its variants significantly outperform the benchmark, especially for low SNRs.
This is mainly because FAST manages to transmit the features with high importance when the CSI is favorable.
In addition, the methods that employ predicted CSI perform worse than those with perfect CSI since the channel predictor is not error-free and a resultant performance loss is inevitable.

\begin{figure}[t]
	\centering
	\includegraphics[width=0.45\textwidth]{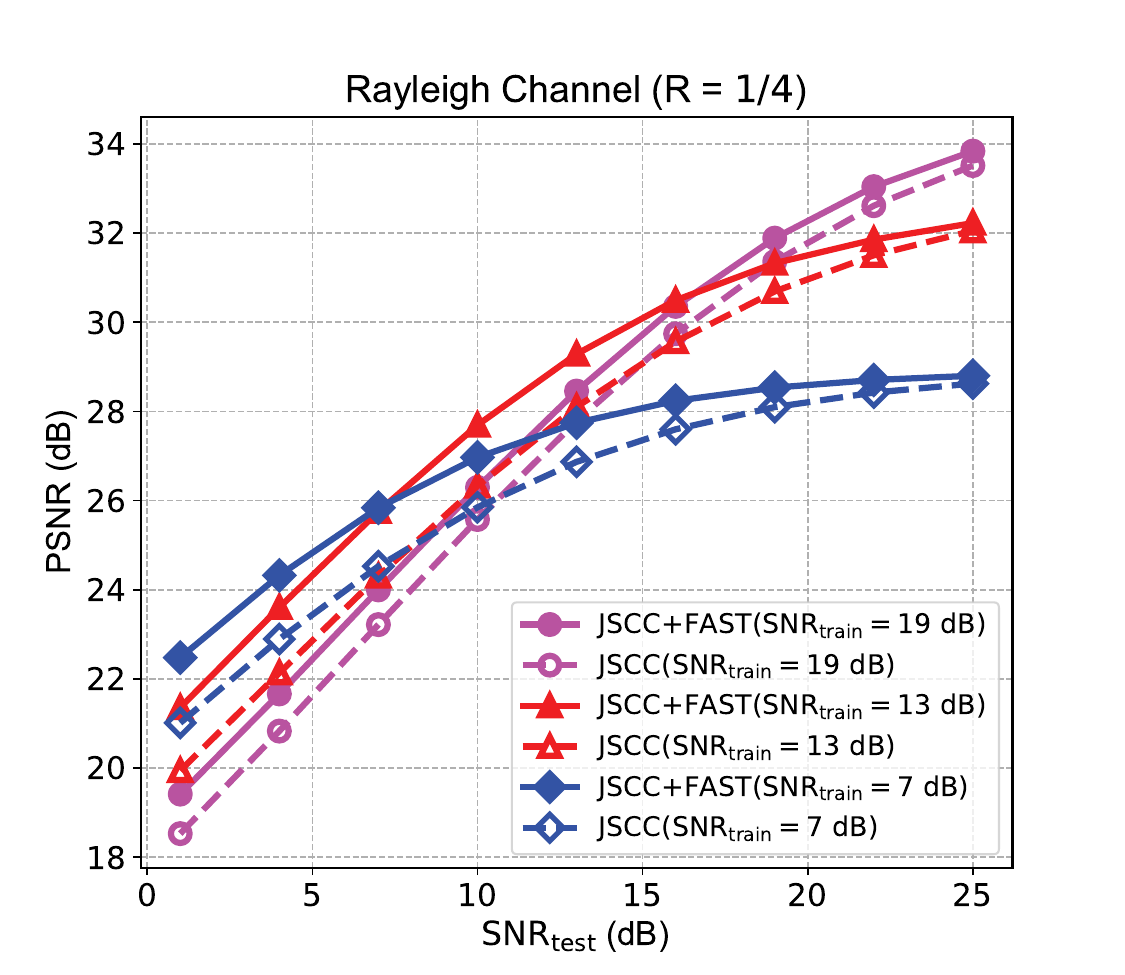}
	\captionsetup{font=footnotesize}
	\caption{Performance of FAST in the time domain with different training SNRs. The basic models are all trained with a SISO Rayleigh channel at $\text{R} = 1/4$. The results are evaluated on the CIFAR-10 dataset by calculating the average PSNR of $10,000$ images.}
	\label{FinalPerform}
\end{figure}

\begin{figure}[t]
	\centering
	\includegraphics[width=0.45\textwidth]{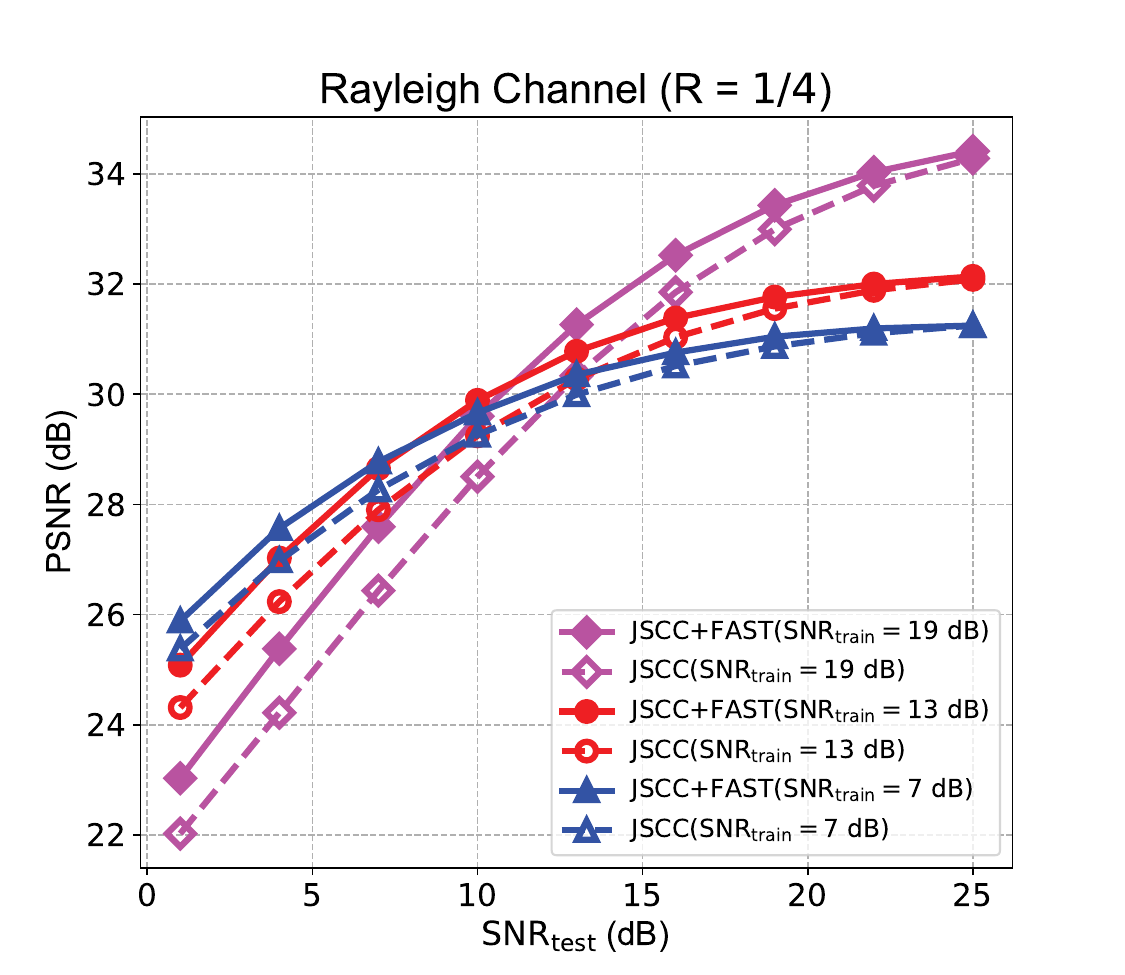}
	\captionsetup{font=footnotesize}
	\caption{Performance of FAST over high-resolution images with different training SNRs. The basic models are all trained on the ImageNet dataset with a SISO Rayleigh channel at $\text{R} = 1/4$. The results are evaluated on the Kodak dataset by calculating the average PSNR of $24$ images.}
	\label{TPerformKodak}
\end{figure}

\begin{figure*}[t]
	\centering
	\includegraphics[width=0.85\textwidth]{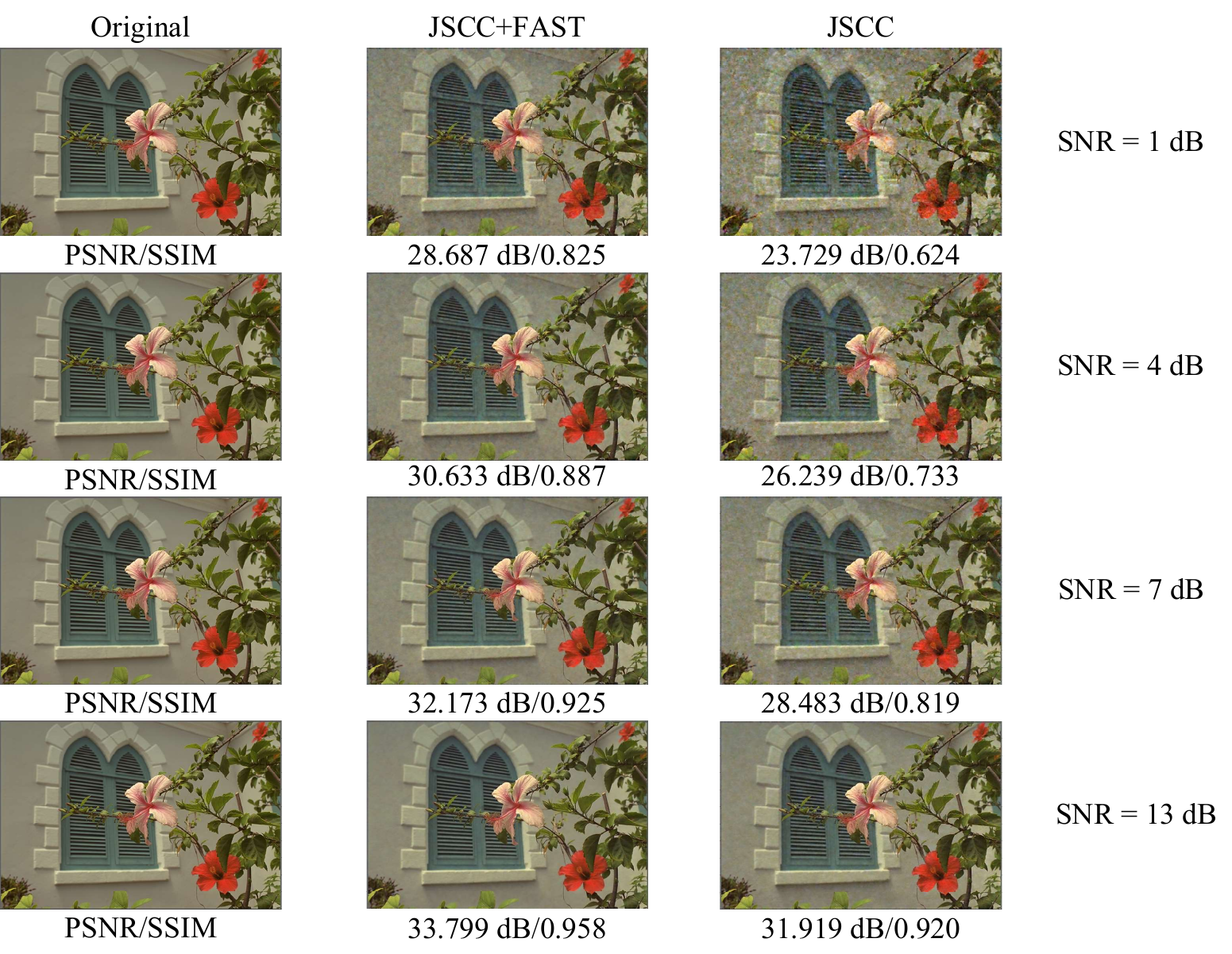}
	\captionsetup{font=footnotesize}
	\caption{Examples of reconstructed images produced by FAST and the benchmark JSCC scheme. Both schemes use the same basic model that is trained on the ImageNet dataset at $\text{R} = 1/4$ and $\text{SNR}_{\text{train}} = 13$ dB. They are evaluated on the Kodak dataset at $\text{R} = 1/4$ and different SNRs. From top to bottom, the rows correspond to testing SNR values of $1$ dB, $4$ dB, $7$ dB, and $13$ dB. We also use a perceptual metric, SSIM, to evaluate image quality.}
	\label{Images}
\end{figure*}

Fig. \ref{FinalPerform} illustrates the performance of the proposed FAST approach for different training SNRs.
We observe that all three models trained at different SNRs with FAST outperform their counterparts without FAST over the entire $\mathrm{SNR_{test}}$ region.
Moreover, the performance gain provided by FAST for models trained at low SNR is larger than that achieved by models trained at high SNR.
This is because FAST mitigates the performance degradation caused by the corruption of important features, especially at low SNRs.
This result exhibits the advantages of the proposed FAST approach in harsh channel conditions.

Fig. \ref{TPerformKodak} exhibits the performance of FAST on high-resolution images.
The basic models are trained using ImageNet and evaluated on the Kodak dataset.
Akin to the outcomes observed on the CIFAR-10 dataset, the models trained with FAST surpass their counterparts across the entire $\mathrm{SNR_{test}}$ range.
This demonstrates that the proposed FAST approach can maintain excellent performance on high-resolution images.
Notably, for high-resolution images, the performance improvement introduced by FAST is more pronounced for models trained at high SNR compared to those trained at low SNR.
This is in contrast to the results observed for low-resolution images in CIFAR-10. 
The discrepancy can be attributed to the presence of larger-sized features in higher resolution images.
Consequently, these features are more exposed to channel noise, rendering them more fragile.
In low SNR environments, these delicate features can be easily affected even after feature allocation.

Fig. \ref{Images} provides a visual comparison of the reconstructed images produced by FAST and the benchmark JSCC approach.
Both schemes are evaluated using the same basic model that had been trained on the ImageNet dataset.
We sample an image in the Kodak dataset and perform reconstructions using FAST and JSCC, respectively.
Each row in the figure corresponds to a different test SNR value, with lower SNR values at the top and progressively higher SNR values toward the bottom.
We also provide PSNR and SSIM scores of each reconstructed image.

\begin{figure}[t]
	\centering
	\subfloat[Precoding-free case.]{\includegraphics[width=0.45\textwidth]{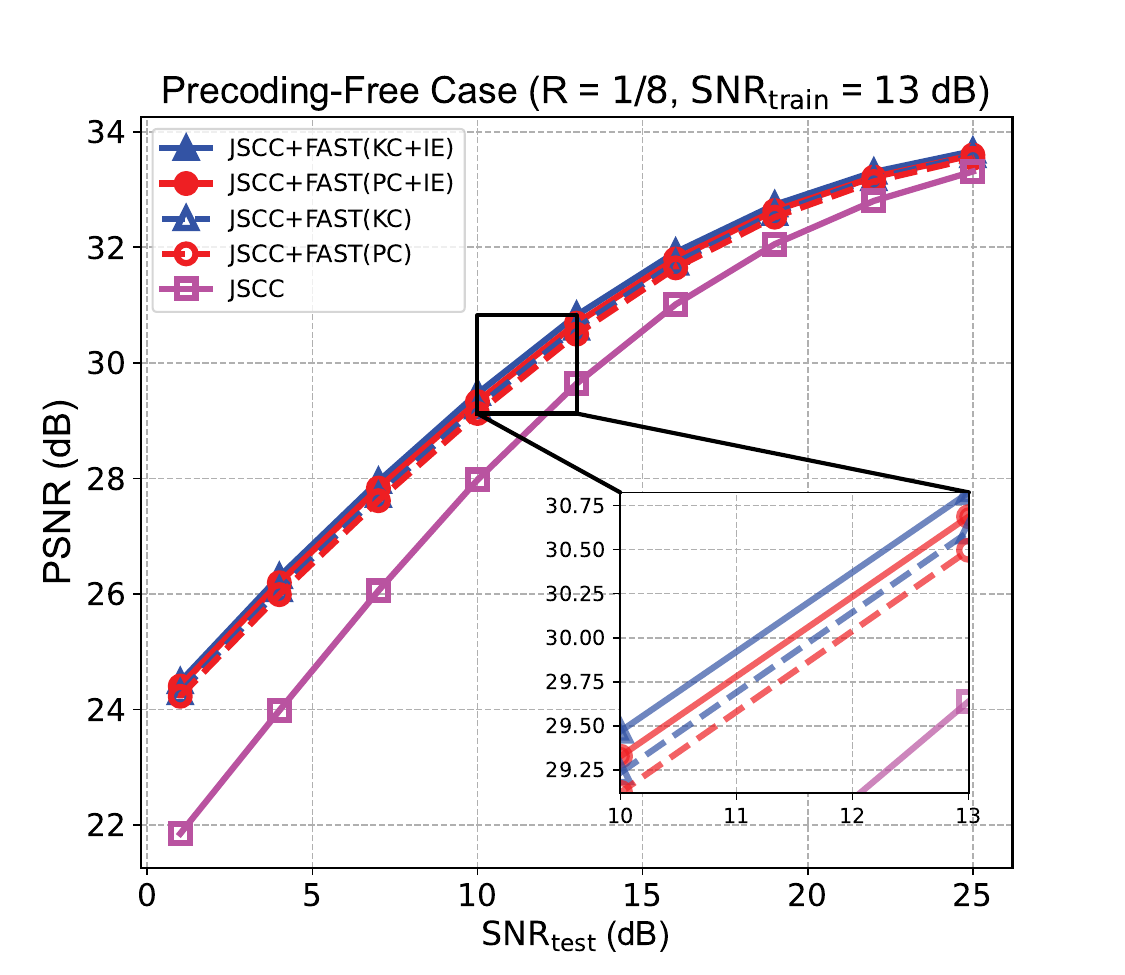}\label{CtrlNonSVD}}
	
	\subfloat[Precoding-based case.]{\includegraphics[width=0.45\textwidth]{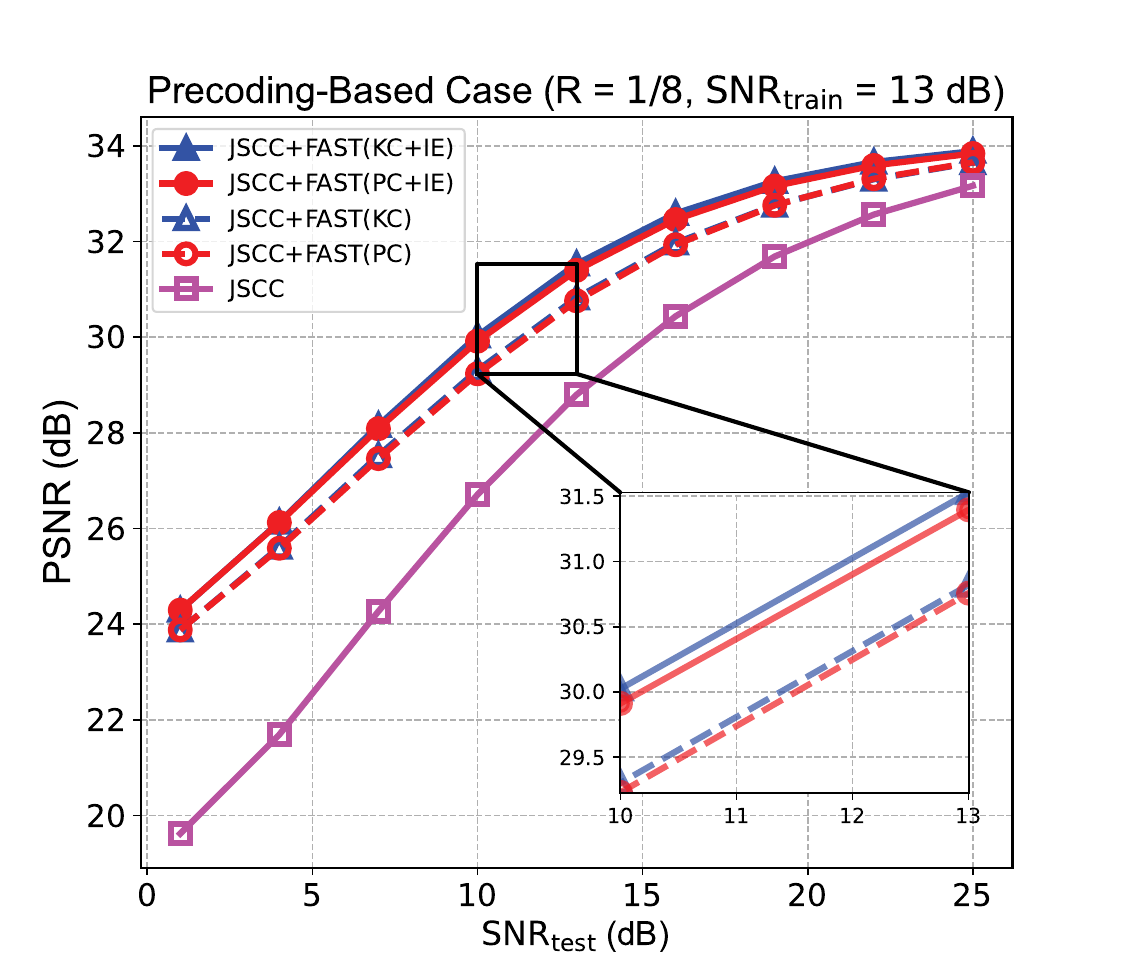}\label{CtrlSVD}}
	\captionsetup{font=footnotesize}
	\caption{Comparison of different schemes in the space-time domain. The basic models are all trained with a SISO Rayleigh channel at $\text{R} = 1/4$ and $\text{SNR}_{\text{train}} = 13$ dB. The schemes are tested with a MIMO Rayleigh channel at $\text{R} = 1/8$. (a) and (b) present the results of the precoding-free and precoding-based cases, respectively. Each point of the curves is obtained by calculating the average PSNR of 10,000 images in the test dataset. }
	\label{STComparison}
\end{figure}

\subsection{Performance of FAST in the Space-Time Domain}
Next, we assess the performance of FAST in the space-time domain.
We have asserted that the feature allocator has an advantage in not necessitating fine-tuning and can be readily applied even when the system configuration, such as the number of antennas, undergoes changes, all while maintaining a performance improvement.
To validate this, we train the basic models with a SISO channel and subsequently evaluate them in a MIMO scenario.
This introduces a substantial mismatch in the system configuration between the training and testing phases.

As with the single-antenna scenario, we simulate five different methods and compare their performance.
Fig. \ref{STComparison} presents the simulation results for both the precoding-free and precoding-based cases.
Notably, we observe that despite the substantial disparity in the system configuration, FAST continues to deliver a remarkable performance enhancement, especially in low SNR environments.
This suggests that FAST can effectively substitute the need for model fine-tuning when the system configuration undergoes changes.
Furthermore, the performance degradation resulting from channel prediction errors is reduced compared to the single-antenna scenario. This is attributed to the introduction of the spatial domain, which offers subchannels in an additional dimension, allowing for shorter predicted CSI sequences and thus mitigating the accumulation of prediction errors.

\begin{figure}[t]
	\centering
	\subfloat[Precoding-free case.]{\includegraphics[width=0.45\textwidth]{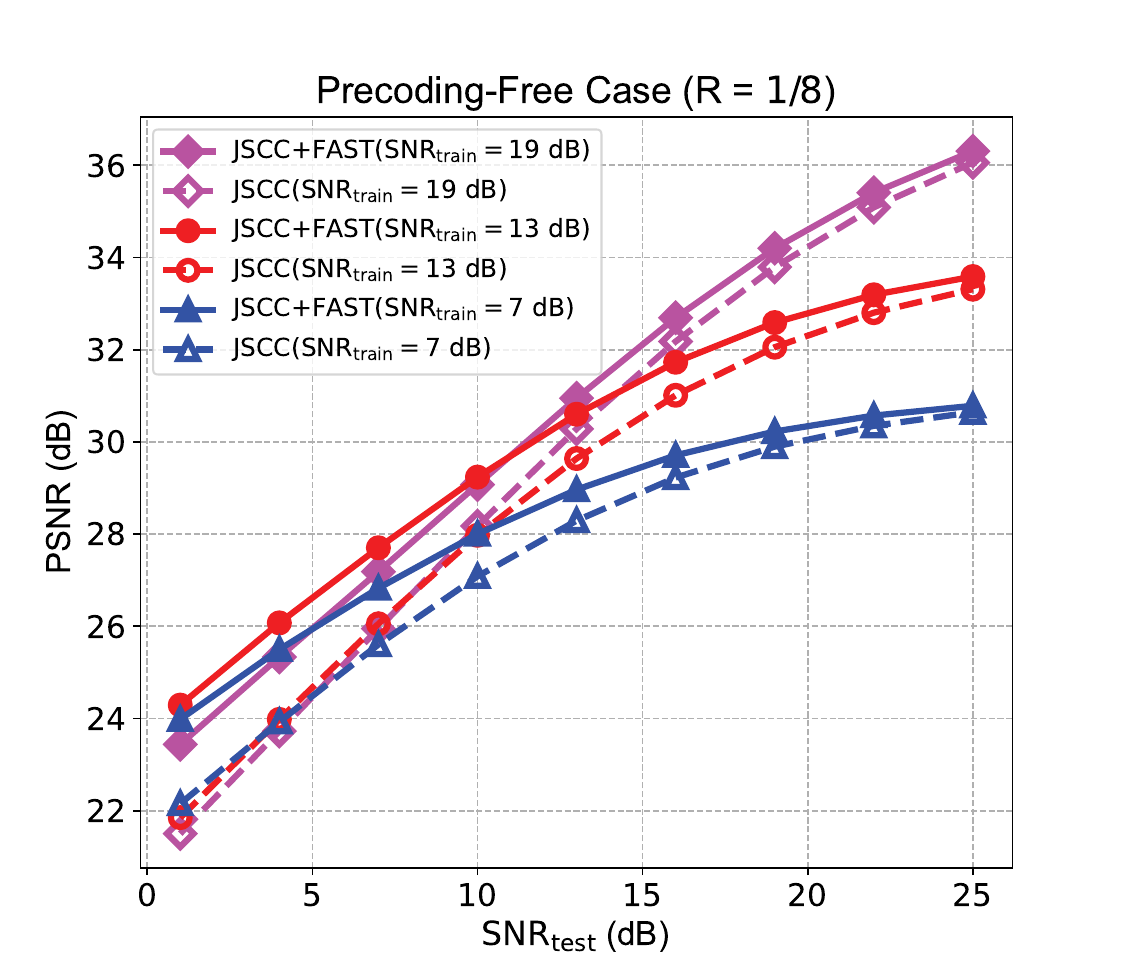}\label{PerformNonSVD}}

	\subfloat[Precoding-based case.]{\includegraphics[width=0.45\textwidth]{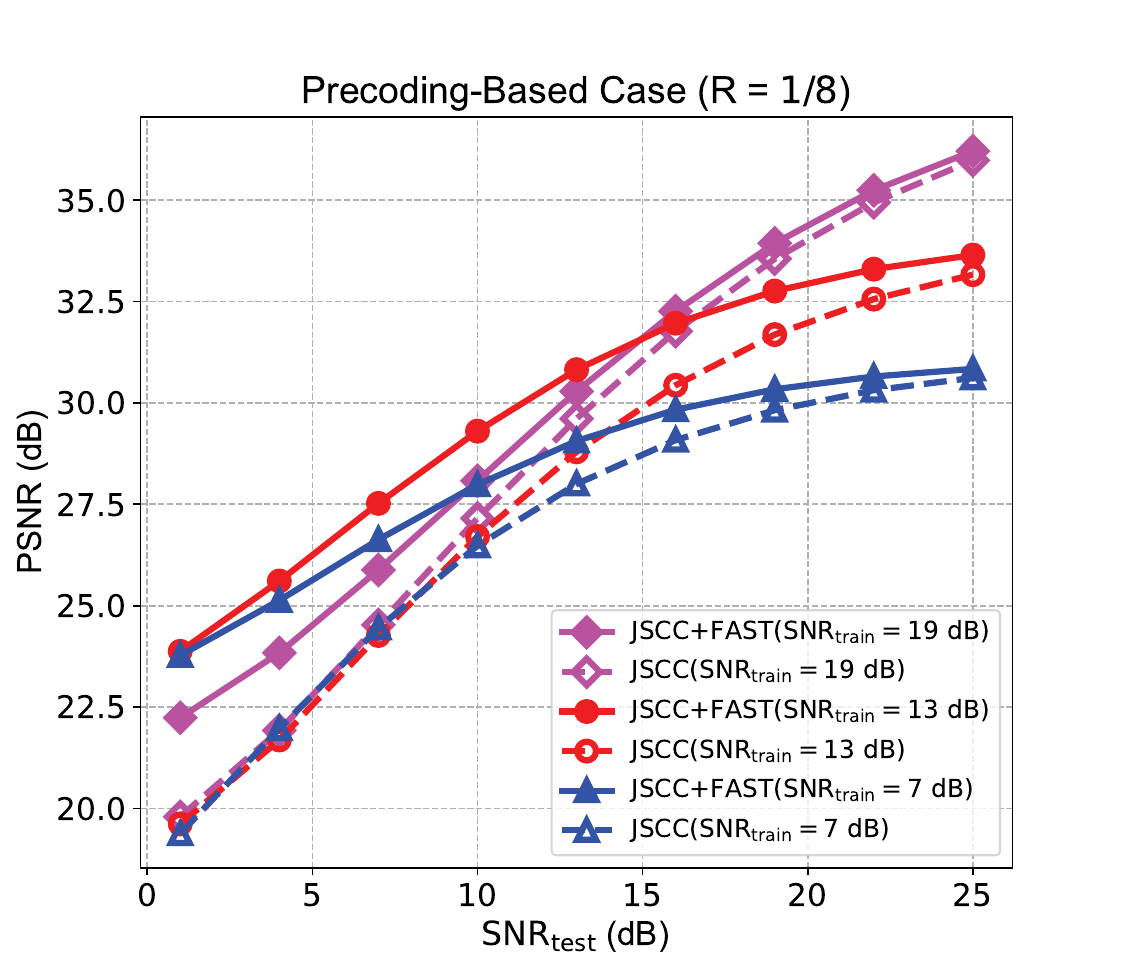}\label{PerformSVD}}
	\captionsetup{font=footnotesize}
	\caption{Performance of FAST in the space-time domain with different training SNRs of the basic models. The training is performed with a SISO Rayleigh channel at $\text{R} = 1/4$ and the evaluation is conducted with a MIMO Rayleigh channel at $\text{R} = 1/8$. (a) and (b) present the results of the precoding-free and precoding-based cases, respectively. The results are evaluated by calculating the average PSNR of 10,000 images in the test dataset .}
	\label{STFinalPerform}
\end{figure}

The performance of FAST in the space-time domain is illustrated in Fig. \ref{STFinalPerform}.
It is apparent that all three models trained at various SNRs demonstrate similar performance in low SNR conditions.
This similarity can be attributed to the configuration mismatch between the training and testing phases.
Moreover, akin to the performance of FAST in the time domain, it maintains its superiority across different training SNRs. Additionally, the precoding-free system consistently offers performance closely aligned with the precoding-based system across various training SNRs.
This further substantiates the practicality of precoding-free systems enhanced by FAST, as it substantially reduces the feedback overhead associated with precoding-based systems while ensuring consistent performance.

\section{Conclusion} \label{Conclusion}
In this paper, we introduced a versatile multi-dimensional feature allocation framework for semantic communication, aimed at enhancing the performance of image transmission.
Our primary goal was to transmit critical features under favorable channel conditions while transmitting less vital ones under lower-quality channel states.
To achieve this, we initially designed an importance evaluator to assess the significance of various features before transmission.
Subsequently, we developed a feature allocator that operates in the time domain, facilitating the allocation of features to appropriate transmission time slots.
Furthermore, we extended the proposed framework to the spatial domain, introducing a space-time domain feature allocator for MIMO systems.
This framework requires no intricate fine-tuning and can be seamlessly deployed, even when the system configuration undergoes changes, all while maintaining notable performance enhancements.
We also addressed both precoding-free and precoding-based MIMO systems, designing distinct feature allocation strategies for each.
Looking ahead, there is the potential for expanding the framework into the frequency domain and integrating it with multi-carrier techniques, which holds promise for further improvements in performance.

\bibliographystyle{IEEEtran}
\bibliography{IEEEabrv,Reference}

\end{document}